%% file: iclr2025_conference.tex
\documentclass{article}
\usepackage{iclr2025_conference,times}
\usepackage[T1]{fontenc}
\usepackage{amsmath,amssymb,amsthm,mathtools}
\usepackage{graphicx,booktabs,array,tabularx,microtype}
\usepackage{url}
\usepackage[hidelinks,breaklinks=true]{hyperref}
\providecommand{\doi}[1]{doi: \begingroup\urlstyle{rm}\href{https://doi.org/#1}{\nolinkurl{#1}}\endgroup}
\newtheorem{theorem}{Theorem}[section]
\newtheorem{proposition}[theorem]{Proposition}

\newcommand{\dd}{\mathrm{d}}

\title{Winning Reviews, Losing Discovery?\\[0.35em]{\large\normalfont The Co-Evolutionary Game of AI Research and Peer Review}}
\hypersetup{pdfauthor={Hongze Liu, Mengru Wang, Zhiyuan Zhang, Yao Liu},pdftitle={Winning Reviews, Losing Discovery? The Co-Evolutionary Game of AI Research and Peer Review}}
\author{Hongze Liu$^{1,*}$ \quad Mengru Wang$^2$ \quad Zhiyuan Zhang$^1$ \quad Yao Liu$^3$\\[3pt]
{\normalfont\small $^1$Department of CSE, SJTU}\\
{\normalfont\small $^2$Faculty of Human Sciences, Waseda University}\\
{\normalfont\small $^3$Center for Human Capital and Labor Market Research, CUFE}\\[3pt]
{\normalfont\small\texttt{seniordriver233@sjtu.edu.cn}\quad\texttt{mengruw@ruri.waseda.jp}}\\
{\normalfont\small\texttt{yukio713@sjtu.edu.cn}\quad\texttt{2025212959@gmail.cufe.edu.cn}}\\
{\normalfont\small $^*$Corresponding author: Hongze Liu}
}
\iclrfinalcopy
\begin{document}
\raggedbottom
\setlength{\parskip}{6pt plus 1pt minus 0.5pt}
\maketitle
\lhead{Winning Reviews, Losing Discovery?}
\rhead{Working manuscript}
\begin{abstract}
Autonomous research systems can produce convincing manuscripts whose scientific value is difficult to verify, while automated reviewers can reward the appearance of completeness. We examine how this gap can become a self-reinforcing threat to the research community. In a two-population evolutionary game, authors choose between acceptance-oriented and exploratory research, and reviewers choose between proxy evaluation and careful assessment. Cheaper acceptance-oriented production can increase review pressure, making proxy assessment more attractive; proxy assessment can in turn strengthen the incentive to produce reviewer-compatible work. We derive conditions under which this feedback sustains a stable regime that rewards publication success while reducing the scientific value of accepted work. An analytical basin boundary explains why the same institutional rules can support either this regime or exploratory research with careful review. Numerical illustrations show how publication counts can conceal scientific loss, how faster adjustment can reinforce existing practices, and why partial reform may fail once the cycle is established. Careful audits and reviewer rewards can reverse the feedback, while temporary support can produce lasting recovery if it moves the community across the basin boundary. The model identifies a warning mechanism rather than forecasting a particular conference: a review system can remain operational and privately rewarding while becoming less effective at supporting discovery. Its institutional implication is to reward independently validated contributions and consequential checking, rather than accelerate paper production and scoring alone.
\end{abstract}
\begin{quote}
\itshape ``We came for the stars, followed the constellations mystery drew, losing sight of the sky.''
\end{quote}
\input{sections/01_introduction}
\input{sections/02_related}
\input{sections/04_model}
\input{sections/08_numerics}
\input{sections/09_discussion}
\input{sections/11_conclusion}
\par
\begingroup\linespread{1.0}\small
\interlinepenalty=10000
\setlength{\bibsep}{1.5pt}
\bibliography{iclr2025_conference}
\bibliographystyle{iclr2025_conference}
\endgroup
\clearpage\appendix
\input{sections/10_appendix}

\end{document}

%% file: sections/01_introduction.tex
\section{Introduction}
\begin{figure}[tb]\centering\includegraphics[width=\linewidth]{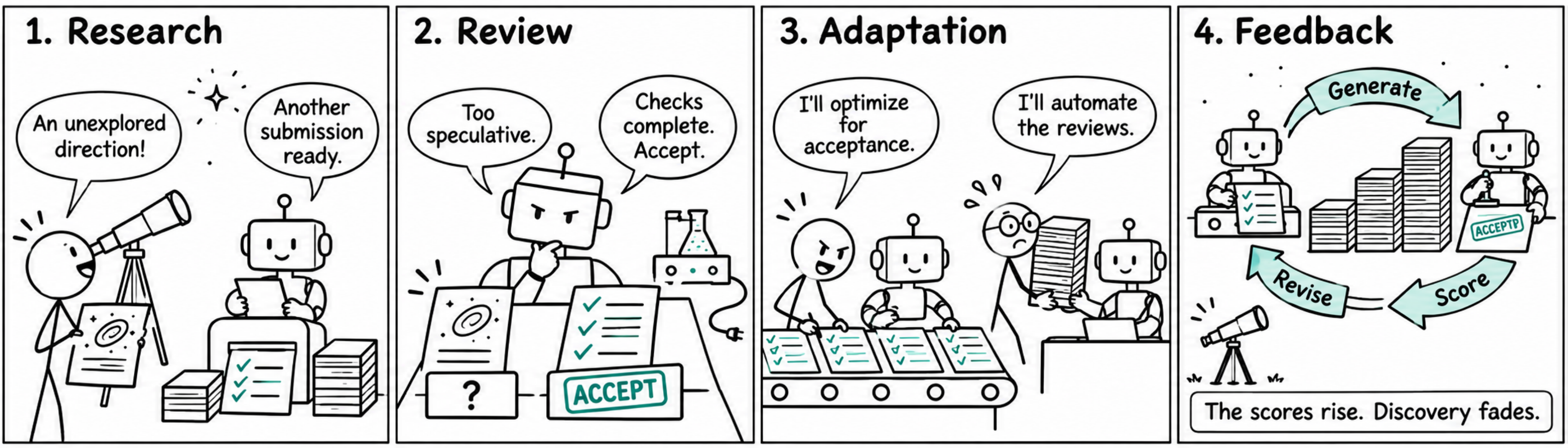}\caption{An illustrative research--review feedback loop. Automated production and proxy evaluation reward reviewer-compatible manuscripts; adaptation can reinforce these practices while crowding out exploration. The comic depicts the mechanism examined in the model.}\label{fig:comic}\end{figure}
The ability to generate a research paper is becoming cheaper than the ability to establish whether it contributes reliable knowledge. Autonomous research systems now connect idea generation, code writing, experimentation, analysis, and manuscript preparation within a single workflow. The AI Scientist additionally evaluates its output through simulated peer review \citep{lu2024scientist}, while its successor introduces agentic tree search to manage experimental exploration \citep{yamada2025scientist}. Such systems move automation beyond assistance with individual tasks toward the production of complete submissions. In workflows informally called ``vibe research,'' researchers can direct repeated cycles of generation and revision without closely examining every intermediate decision. This creates an institutional vulnerability: a system may produce a convincing manuscript without reliable findings, while an automated reviewer rewards the manuscript without detecting that failure. When the same class of tools supplies both research and its evaluation, inexpensive production can expand faster than the capacity for independent scientific judgment. The central danger is that this separation becomes self-reinforcing: the community rewards successful submissions while losing the incentives and evaluative attention that discovery requires.

Independent evaluations provide concrete evidence that autonomous research can produce papers of poor scientific quality. In their study of Sakana's AI Scientist, \citet{beel2025evaluation} report that five of twelve proposed experiments failed because of coding errors. Successful execution did not ensure sound research: four of the seven resulting manuscripts contained incorrect or hallucinated numerical results. The system also misclassified established techniques as novel and produced manuscripts with weak literature coverage, incomplete sections, and missing or misplaced figures. These problems concern the relationship between claims, implementation, and evidence, rather than writing style alone. A subsequent audit of two research systems identifies benchmark-selection vulnerabilities, data leakage, metric misuse, and post hoc selection that can remain largely hidden in the final manuscript \citep{luo2026pitfalls}. In its constructed audit dataset, manuscript-only detection accuracy was 53.7\%, rising to 81.5\% when the auditor received workflow artifacts. Complementing these system studies, SciIntegrity-Bench reports a 34.2\% combined failure-and-flawed-output rate across 231 runs in deliberately constructed research-integrity dilemmas \citep{yang2026integrity}. The studies examine different settings, but identify a common failure: completing a plausible research artifact can take precedence over establishing that its conclusions follow from what was actually done.

Automated reviewing can fail precisely where this distinction matters. In the adversarial BadScientist study, agents generate fabricated papers without conducting real experiments, yet achieve acceptance rates of up to 82\% from the tested LLM review systems \citep{jiang2026badscientist}. Some reviewers flag integrity concerns while still assigning acceptance-level scores, demonstrating that producing criticism does not ensure that criticism governs the recommendation. Content-preserving rhetorical revisions can also change AI review scores, providing a route to evaluation gains without corresponding scientific improvement \citep{li2026rhetoric}. At the same time, review attention may be poorly allocated: off-the-shelf LLMs emphasize technical validity while overlooking novelty relative to human experts \citep{shin2025blindspots}. In a study of 139 Austrian grant proposals, those rated as more novel by humans are more likely to be selected by humans but not by GPT-4o \citep{machado2026novelty}. These findings reveal two compatible risks. Automated evaluation may reward an unsound paper whose presentation resembles rigorous work, while penalizing an uncertain proposal whose scientific promise requires contextual judgment. Research optimized against such evaluations may consequently become increasingly defensive, answering familiar objections and supplying recognizable experimental detail while avoiding questions whose importance is difficult to demonstrate in advance.

The consequence can extend beyond individual bad papers because authors and reviewers respond to the environment created by one another. Rewards tied to observable performance can displace valuable activities that are harder to measure \citep{holmstrom1991}, allowing publication incentives to select practices that advance careers more than discovery \citep{smaldino2016}. Exploratory proposals already face recognition difficulties in human evaluation \citep{boudreau2016}; automation can make adjustment to standardized assessment much faster. Authors who obtain favorable scores at low production cost have an incentive to repeat that strategy. If the resulting submissions increase review load, reviewers gain a stronger incentive to delegate assessment, making the same acceptance signals more influential. Evidence of LLM-associated conference reviewing establishes that automation has already entered this evaluative setting \citep{liang2024}. ``AI helping AI'' then describes a feedback mechanism: production adapts to the evaluator's weaknesses, and evaluation becomes cheaper without becoming more discriminating. A conference can continue processing submissions and issuing decisions even as its rewards increasingly favor reviewer-compatible output over scientific progress.

We formalize this mechanism through a two-population evolutionary game linking research choice with evaluation effort. Authors choose acceptance-oriented or exploratory research; reviewers choose proxy evaluation or careful assessment. These are strategies that humans and machines can both adopt, with author payoffs determined by acceptance rewards and production costs, and reviewer payoffs by effort savings, workload, and the benefit of careful judgment. Existing work examines verification incentives \citep{you2026} and the wider production--evaluation arms race \citep{wang2026}; our contribution is to characterize their coupled adaptation, equilibrium selection, and recovery conditions. When each population sufficiently increases the other's return to proxy-oriented behavior, the model admits two attracting configurations separated by a saddle's stable manifold. Independently defined scientific value then determines whether the privately successful configuration is socially inferior. This distinction makes the warning precise: the community can settle into a privately rewarding cycle that maintains publication activity while weakening discovery. Individual success then sustains the collective loss instead of correcting it. We analyze how careful audits and reviewer rewards alter these incentives and use reproducible numerical illustrations to examine transitions between regimes. Our contributions are:
\begin{itemize}
\item We develop a two-population evolutionary model of research production and peer review, formalizing how acceptance incentives and workload-dependent evaluation costs jointly shape the adoption of research and review strategies.
\item We derive conditions for bistability and characterize the boundary between basins of attraction analytically, establishing how initial conditions and relative adaptation speeds influence equilibrium selection.
\item We introduce an independent scientific-value criterion and derive conditions under which a stable acceptance-oriented equilibrium yields less scientific value, distinguishing publication volume from contribution quality and accounting for differences in production rates.
\item We derive sufficient intervention thresholds for careful audits and reviewer rewards, and distinguish equilibrium destabilization from transitions between basins. Reproducible numerical analyses examine intervention timing and the robustness of the resulting dynamics.
\end{itemize}

%% file: sections/02_related.tex
\section{Related work}
The warning developed here connects a familiar incentive problem with a new production technology: research can be selected for its ability to satisfy evaluation while contributing less to discovery. We bring together work on scientific incentives, direct evaluations of automated research and review, and institutional feedback. Their connection motivates studying how authors and reviewers sustain one another's practices.

\subsection{Scientific incentives and exploration}
Multitask incentive theory explains how rewards for measurable performance divert effort from less measurable contributions \citep{holmstrom1991}. In science, publication-based selection can propagate weaker research practices even without deliberate misconduct \citep{smaldino2016}. Exploratory work is particularly exposed because its importance can be difficult to establish before its implications are understood: randomized evaluator--proposal assignments reveal lower scores for highly novel proposals and effects of intellectual distance \citep{boudreau2016}. Automation increases the speed at which researchers can adapt to these rewards. The resulting concern extends beyond erroneous individual papers to the selection of questions, methods, and practices that a community continues to support. The epistemic risks described by \citet{messeri2024} reinforce the need to distinguish scientific understanding from increased output.

\subsection{Automated research and evaluation}
End-to-end systems now combine idea generation, experimentation, writing, and automated assessment \citep{lu2024scientist,yamada2025scientist}. Independent evaluations document execution failures, incorrect numerical claims, and hidden methodological problems \citep{beel2025evaluation,luo2026pitfalls}; integrity benchmarks additionally expose failures under completion pressure \citep{yang2026integrity}. On the evaluation side, BadScientist demonstrates acceptance-level scoring of fabricated papers in adversarial tests \citep{jiang2026badscientist}, while rhetorical rewriting changes scores without changing reported content \citep{li2026rhetoric}. Review-focus comparisons identify greater emphasis on technical validity and less attention to novelty \citep{shin2025blindspots}, alongside under-selection of human-rated novel proposals in a grant study \citep{machado2026novelty}. These findings motivate a quality--score gap through which automated production can exploit automated evaluation. Corpus evidence documents assistance in conference reviewing \citep{liang2024}; the policy experiment of \citet{kim2026} and quality-adjusted analysis of \citet{sharma2026} further distinguish this behavioral mechanism from tool-use labels or inherent AI-authorship favoritism.

\subsection{Feedback and institutional design}
Verification-pressure models explain how scarce checking weakens incentives for truth-oriented effort \citep{you2026}, and the scholarly arms-race account describes wider production--evaluation co-adaptation \citep{wang2026}. We model author and reviewer choices as coupled populations, identifying when harmful practices become stable and when coordinated intervention can reverse them. Reviewer recruitment \citep{stelmakh2020}, review-quality assessment \citep{goldberg2024}, Registered Reports \citep{chambers2013,chambers2022}, and transparency standards \citep{nosek2015} provide institutional precedents for changing capacity, recognition, and access to evidence. Our analysis connects such changes to equilibrium selection: temporary support can have a lasting effect when ordinary incentives resume from a different basin. Standard evolutionary-game methods \citep{hofbauer1998} thus provide a way to analyze institutional persistence and recovery, while a separate scientific-value objective identifies the collective loss that private rewards can conceal.

%% file: sections/04_model.tex
\section{The Author--Reviewer Evolutionary Game}\label{sec:model}
The central warning is that authors and reviewers can sustain practices that damage discovery even when each responds rationally to private rewards. We model how inexpensive acceptance-oriented production changes review pressure, how proxy assessment changes the reward for exploration, and when the resulting feedback becomes stable. An independent scientific-value criterion identifies the collective loss that this private stability can conceal.

The empirical setting combines substantial evaluation demand with increasing access to low-cost assessment. The ICLR 2026 retrospective reports 19,525 valid submissions and 76,139 reviews \citep{iclr2026}; corpus studies document LLM-associated review writing \citep{liang2024}. Quality-adjusted comparisons \citep{sharma2026} and the ICML policy experiment \citep{kim2026} distinguish actual judgment from authorship and policy labels. The model therefore represents the relevant practices directly: acceptance incentives, checking effort, and workload determine the feedback, rather than tool use alone.

The two empirical links are the change in research rewards under different evaluation protocols, represented by $\beta$, and the change in reviewers' effort incentives under production pressure, represented by $\kappa$. Their interaction determines whether cheaper production weakens the evaluation needed to discipline it. Appendix~\ref{app:evidence} records source populations, protocol differences, and the measurements needed to estimate these links. The numerical study uses declared illustrative inputs.

\subsection{Strategies and incentives}
Let $x\in[0,1]$ be the share of authors using acceptance-oriented research $A$, with the remaining share using exploratory research $T$. Let $y\in[0,1]$ be the share of evaluations using proxy assessment $M$, with the remainder using careful assessment $C$. Acceptance-oriented work is organized around expected evaluation rewards; exploratory work accepts greater uncertainty about recognition in pursuing a contribution. Proxy assessment relies on inexpensive signals of merit, while careful assessment invests in understanding and checking consequential claims. These distinctions concern practices rather than author identities: automated tools can support exploration and verification, and human participants can rely on superficial signals.

The baseline unit is one submission opportunity per author during an adjustment period. Let $p_i^j\in[0,1]$ denote acceptance probability for strategy $i$ under protocol $j$, $B>0$ the private benefit of acceptance, and $c_i\geq0$ research cost per opportunity. Random exposure to the current reviewer population gives
\begin{equation}\label{eq:author}
 p_i(y)=(1-y)p_i^C+yp_i^M,\qquad U_i(y)=Bp_i(y)-c_i.
\end{equation}
The mixture summarizes a protocol rather than a literal average of committee scores. Fixed acceptance capacity and multiple-reviewer decision rules can introduce additional dependencies; the baseline deliberately isolates the two-population interaction before examining those extensions.

Writing $\Delta_C=p_A^C-p_T^C$ and $\Delta_M=p_A^M-p_T^M$, the relative author payoff is
\begin{align}
 D_A(y)&=U_A(y)-U_T(y)=\alpha+\beta y,\label{eq:DA}\\
 \alpha&=c_T-c_A+B\Delta_C,\qquad
 \beta=B(\Delta_M-\Delta_C).\label{eq:ab}
\end{align}
The author compares the extra chance of acceptance with the extra cost of producing the work. The intercept $\alpha$ gives this comparison when all assessment is careful; $\beta$ describes how it changes as proxy assessment spreads. When $\beta>0$, authors have progressively less reason to bear the uncertainty and effort of exploration. This shift can occur even if careful reviewers still prefer exploratory submissions: a modest acceptance disadvantage may be outweighed by producing acceptance-oriented work more cheaply.

For reviewers, let load be $L(x)=L_0+\ell x$, nonnegative throughout the state space. Let proxy assessment save $\delta t\geq0$ payoff units per load unit, careful assessment provide a privately valued accuracy benefit $\eta\Delta q$, and proxy assessment incur an additional cost $d$. A common completion reward cancels, leaving
\begin{equation}\label{eq:reviewer}
 D_R(x)=\delta t(L_0+\ell x)-\eta\Delta q-d=\gamma+\kappa x,
\end{equation}
where $\gamma=\delta tL_0-\eta\Delta q-d$ and $\kappa=\delta t\ell$. A reviewer gives up the time saved by proxy assessment only when careful checking is sufficiently worthwhile to them. If acceptance-oriented production increases the number or checking burden of submissions, that sacrifice grows and proxy assessment becomes more attractive. Expanding reviewer supply or making claims easier to verify can interrupt this link, so $\ell$ need not be positive. If careful assessment's private accuracy benefit changes with composition, its marginal effect also belongs in $\kappa$.

Higher-payoff practices increase their population shares according to
\begin{equation}\label{eq:dynamics}
 \dot x=x(1-x)(\alpha+\beta y),\qquad
 \dot y=\rho y(1-y)(\gamma+\kappa x),\quad\rho>0.
\end{equation}
The parameter $\rho$ combines relative reviewer adjustment speed and payoff scale. Model time has no calendar interpretation. Authors and reviewers may be the same people, but the equations represent their roles without modeling within-person correlations. Absent strategies remain absent under this imitation baseline; convergence statements below concern interior initial states. Small switching rates relax this boundary property in the robustness study.

\begin{figure}[tb]\centering\includegraphics[width=\linewidth]{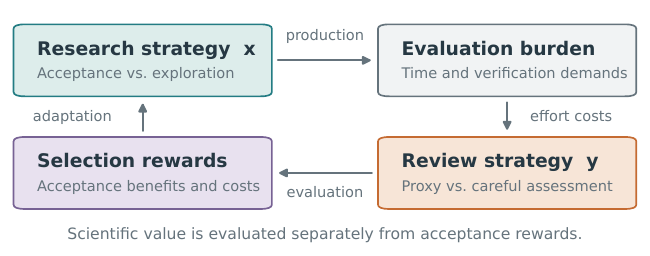}\caption{The coupled incentive mechanism. Production changes evaluation demands; review effort changes the acceptance rewards faced by authors. Scientific value is assessed separately from private rewards.}\label{fig:mechanism}\end{figure}

\subsection{Equilibrium selection and scientific value}\label{sec:theory}\label{sec:welfare}
Each corner is an equilibrium. When $\beta\kappa\neq0$, an interior equilibrium exists exactly when
\begin{equation}\label{eq:thresholds}
 x^*=-\gamma/\kappa\in(0,1),\qquad y^*=-\alpha/\beta\in(0,1).
\end{equation}
For positive cross-effects, $y^*$ is the proxy-review share above which acceptance-oriented work pays better, and $x^*$ is the author share above which reviewers prefer proxy assessment. Below both thresholds, exploration and careful review reinforce each other; above both, acceptance-oriented production and proxy review do so. When only one threshold is exceeded, the two populations initially move in opposite directions. Their subsequent interaction determines which practice becomes established. Two stable outcomes require both payoff differences to switch sign within the admissible range.

\begin{theorem}[Coordination and path dependence]\label{thm:bistable}
Suppose $\beta,\kappa,\rho>0$ and
\begin{equation}\label{eq:bistability}
 \alpha<0<\alpha+\beta,\qquad\gamma<0<\gamma+\kappa.
\end{equation}
Then $(0,0)$ and $(1,1)$ are locally asymptotically stable, the other corners are repelling, and $(x^*,y^*)$ is a saddle. Its stable manifold separates two basins; every interior trajectory outside that manifold converges to one of the attracting corners.
\end{theorem}
The corner eigenvalues are $(\alpha,\rho\gamma)$ at $(0,0)$ and $(-\alpha-\beta,-\rho[\gamma+\kappa])$ at $(1,1)$. At the interior equilibrium the Jacobian has zero diagonal and
\begin{equation}\label{eq:interioreigen}
 \lambda^2=\rho\beta\kappa x^*(1-x^*)y^*(1-y^*)>0,
\end{equation}
giving one attracting and one repelling direction. To characterize the boundary globally, define
\begin{equation}\label{eq:FG}
\begin{aligned}
 F(x)&=\gamma\log x-(\gamma+\kappa)\log(1-x),\\
 G(y)&=\alpha\log y-(\alpha+\beta)\log(1-y).
\end{aligned}
\end{equation}
Direct differentiation gives the conserved quantity
\begin{equation}\label{eq:firstintegral}
 H(x,y)=F(x)-G(y)/\rho.
\end{equation}
The stable separatrix is the decreasing branch through the saddle of
\begin{equation}\label{eq:separatrix}
 G(y)-G(y^*)=\rho[F(x)-F(x^*)].
\end{equation}
The increasing branch is the unstable manifold, so the entire level set must not be treated as a basin boundary. Within the lower-left threshold rectangle both coordinates decrease toward the careful corner; within the upper-right rectangle both increase toward the proxy corner. The stable branch separates which rectangle is eventually entered from the off-diagonal regions. Appendix~\ref{app:proofs} supplies the complete argument.

This geometry explains why a community can remain in an undesirable regime without any participant choosing that collective outcome. When most evaluation is careful, exploratory work receives enough recognition to sustain it, and manageable load supports careful review. When proxy assessment is prevalent, switching research strategy alone may bring little reward, while switching review practice alone requires more effort. \textbf{Each side waits for an environment the other side is not rewarded for creating.} The basin boundary identifies how much coordinated change is needed to escape this situation. Changing $\rho$ changes who adjusts first and can therefore change the eventual outcome even though neither population's payoff threshold moves.

Cost changes act on different parts of this geometry. Holding evaluation probabilities fixed, increasing $c_T-c_A$ lowers $y^*$ at rate $-1/\beta$; increasing $\gamma$ lowers $x^*$ at rate $-1/\kappa$. The careful corner loses strict local stability at $\alpha=0$ or $\gamma=0$. A population can also cross a basin boundary before either equilibrium disappears. The latter changes the trajectory's destination, whereas a stability boundary changes the game itself. Degenerate cases and alternative feedback signs are treated in Appendix~\ref{app:extensions}.

The local classification also clarifies what the coordination result excludes. The square is invariant because the vector field has zero normal component on each edge, and polynomial continuity gives a unique global forward solution for any admissible state. At the opposite corners, the eigenvalues are $(-\alpha,\rho[\gamma+\kappa])$ at $(1,0)$ and $(\alpha+\beta,-\rho\gamma)$ at $(0,1)$. Under the theorem's strict signs both corners repel, leaving the stable saddle branch as the interior boundary between outcomes. At an endpoint equality an edge can become stationary, so strict stability statements cannot be extended through zero eigenvalues. These boundary cases matter for interpreting policies close to their threshold.

The reinforcing signs identify the mechanism that puts the community at risk. If $\beta=0$, authors' private choices no longer respond to review composition; if $\kappa=0$, author composition no longer changes reviewers' effort incentives. Opposite signs can produce a center rather than a saddle at an interior crossing. These cases show where the harmful reinforcement is interrupted. The warning rests on the two populations strengthening each other's incentive to rely on acceptance signals, not on automation inevitably producing a single outcome.

Scientific performance requires an objective beyond private payoff. Let $v_i\geq0$ be expected independently assessed scientific value conditional on acceptance under strategy $i$, with a declared horizon and unit. Accepted quantity and value per submission opportunity are
\begin{align}
 Q(x,y)&=xp_A(y)+(1-x)p_T(y),\label{eq:quantity}\\
 W(x,y)&=xp_A(y)v_A+(1-x)p_T(y)v_T.\label{eq:welfare}
\end{align}
Average accepted value is $W/Q$ when $Q>0$. The proxy corner is scientifically inferior under this measure if and only if
\begin{equation}\label{eq:bad}
 p_A^Mv_A<p_T^Cv_T.
\end{equation}
The comparison identifies a form of institutional damage that publication statistics can conceal. A venue could accept the same number of papers under both regimes while contributing less because its accepted work is less reliable or consequential. Participants can continue receiving professional rewards even as the scientific return declines. Detecting that decline requires independent assessment of contribution: the values $v_i$ cannot simply be replaced by reviewer scores or novelty labels. Reviewer-dependent values $v_i^j$ allow careful assessment to improve selection within each research strategy without changing the private behavioral equations. The warning concerns the separation of private success from scientific value, which the stability calculation alone cannot reveal.

The baseline assigns the same accepted conditional value to a strategy across protocols. A more selective reviewer might accept a different subset even at the same acceptance rate. Allowing protocol-dependent values gives
\begin{equation}\label{eq:selectivevalue}
\begin{aligned}
W(x,y)={}&x[(1-y)p_A^Cv_A^C+yp_A^Mv_A^M]\\
&+(1-x)[(1-y)p_T^Cv_T^C+yp_T^Mv_T^M].
\end{aligned}
\end{equation}
The corner comparison then uses $p_A^Mv_A^M$ and $p_T^Cv_T^C$. This change preserves the behavioral analysis when value does not enter private payoffs, while exposing another way that evaluation can improve science: selecting more valid or consequential work within the same author category. It also prevents the model from treating equal acceptance rates as evidence of equal selection quality.

If strategy $i$ generates $n_i$ submissions per researcher-period, consistency requires changing both private utility to $n_i[Bp_i(y)-c_i]$ and scientific output to $xn_Ap_A(y)v_A+(1-x)n_Tp_T(y)v_T$. The corresponding author intercept and slope also change. Higher throughput can therefore alter both behavior and the welfare ordering. A fixed acceptance cap makes probabilities depend on population composition, invalidating the baseline's affine author gap. These extensions, derived in Appendix~\ref{app:extensions}, are tested numerically rather than assumed to preserve the coordination theorem.

For the throughput extension, the author payoff remains affine in review composition but has different coefficients:
\begin{equation}
\begin{aligned}
\alpha_n&=n_A(Bp_A^C-c_A)-n_T(Bp_T^C-c_T),\\
\beta_n&=B[n_A(p_A^M-p_A^C)-n_T(p_T^M-p_T^C)].
\end{aligned}
\end{equation}
The scientific comparison becomes $n_Ap_A^Mv_A<n_Tp_T^Cv_T$. Increasing output can thus compensate for lower conditional value while simultaneously strengthening the private incentive to adopt $A$. The rate experiment holds reduced reviewer coefficients fixed to isolate this direct author effect; a fully linked institutional model would also connect total production to load and reviewer supply. Those distinctions make the units of an empirical test important: time per paper, submissions per researcher, and verification burden per reviewer are not interchangeable measurements.

A capacity constraint introduces a different departure. Our sensitivity case proportionally thins acceptance probabilities when unconstrained demand exceeds a cap. It guarantees a fixed accepted-output limit, but does not model merit ranking. Since the thinning multiplier depends on the author population, the author gap now depends on both coordinates and the affine separatrix no longer applies. The numerical comparison tests this alternative directly rather than treating the baseline theorem as a general result about competitive selection.

\subsection{Institutional interventions}\label{sec:policy}
An institution can change the evaluation authors encounter or the incentive to supply careful review. Let a fraction $a\in[0,1]$ receive externally resourced careful assessment that determines the outcome for those opportunities. Effective proxy exposure becomes $z=(1-a)y$, giving
\begin{equation}\label{eq:audit}
 D_A^a(y)=\alpha+\beta(1-a)y.
\end{equation}
The voluntary reviewer state remains $y$; a population with $y=1$ can still receive assigned careful assessments. Holding reviewer incentives fixed assumes the institution supplies this capacity rather than drawing it from the same constrained volunteers.

Under Theorem~\ref{thm:bistable}, the audit threshold is
\begin{equation}\label{eq:athreshold}
 a_c=1+\alpha/\beta\in(0,1).
\end{equation}
For $a>a_c$, authors prefer exploration even when every voluntary reviewer uses proxy assessment, so $x\to0$. Assigned careful evaluations make it too difficult to rely on proxy-compatible acceptance signals alone. As authors change strategy, the pressure favoring proxy review declines, eventually giving $y\to0$. Below this threshold the proxy corner remains locally attracting, although audits may still redirect particular initial states. An already careful environment gains no additional protection against author invasion because $D_A^a(0)=\alpha$. Equality requires separate boundary analysis.

A reward $s\geq0$ for careful reviewing instead changes the reviewer gap to
\begin{equation}\label{eq:incentive}
 D_R^s(x)=\gamma+\kappa x-s.
\end{equation}
If $s>s_c=\gamma+\kappa$, the reward compensates reviewers for careful assessment even under the strongest modeled pressure to use proxies. Review practice then changes first: $y\to0$, after which exploration becomes privately preferable and $x\to0$. Audits and rewards thus approach the same feedback loop from different sides. Smaller rewards move the reviewer threshold to $(s-\gamma)/\kappa$ but leave proxy-corner attraction possible. These sufficient results concern permanent policies and interior states; adjustment can be slow near equality. Appendix~\ref{app:proofs} gives the proofs.

A temporary policy can have a lasting effect if it changes ordinary incentives before support ends. Once enough reviewers assess carefully and enough authors find exploration worthwhile, each population can sustain the other's change without continued assistance. Formally, baseline dynamics resume at time $T$ from the release state: recovery persists if that state is in the careful basin and reverses if it remains in the proxy basin. Ending support too early therefore permits a return to the original regime. The relevant stopping criterion is the state reached, not time or expenditure alone. Whether this recovery is worth its cost also depends on the selected work's value, verification resources, and reviewer opportunity costs.

A useful policy comparison separates gross accepted value from real resource use. Under assigned audits, actual careful exposure is $a+(1-a)(1-y)$ and proxy exposure is $(1-a)y$. With unit checking costs $k_C,k_M$ and a declared conversion $\lambda$ into scientific-value units, one possible objective is
\begin{equation}
J_a=W(x,(1-a)y)-\lambda\{[a+(1-a)(1-y)]k_C+(1-a)yk_M\}.
\end{equation}
A careful-review reward requires budget expenditure $s(1-y)$ per normalized review period, with social cost determined by financing and opportunity costs. The numerical panels report gross value; choosing a policy also requires valuing the resources used. The institutional task is to restore consequential checking and the reward for discovery at a sustainable cost, rather than maximize audit coverage or subsidy payments themselves.

Audits and rewards can also operate together. The proxy-corner eigenvalues become $-[\alpha+\beta(1-a)]$ and $-\rho[\gamma+\kappa-s]$. Either sufficiently strong lever removes strict attraction. Two subthreshold levers can still redirect a specified initial population by shifting the basin boundary, although no such improvement is guaranteed for all initial states. The same principle explains the temporary-policy test: the release state, rather than the policy label or its expenditure alone, determines persistence under the original incentives.

%% file: sections/08_numerics.tex
\section{Numerical results}\label{sec:numerics}
The numerical results show how the research--review cycle can become established, conceal a loss of scientific value, and resist partial reform. We first examine the baseline feedback and interventions, then identify which changes preserve or interrupt it. Parameters are illustrative; full specifications and diagnostics appear in Appendix~\ref{app:numerical}.

\input{figures/generated_results}

\subsection{Baseline behavior and intervention effects}
We use the following normalized parameters to represent a setting where acceptance-oriented work is cheaper to produce and receives an advantage under proxy review:
\begin{equation}\label{eq:baseline}
\begin{gathered}
 B=1,\quad c_A=0.1,\quad c_T=0.2,\\
 p_A^C=0.45,\quad p_T^C=0.65,\quad
 p_A^M=0.65,\quad p_T^M=0.45,\\
 \gamma=-0.2,\quad\kappa=0.4,\quad\rho=1,
 \quad v_A=0.6,\quad v_T=1.
\end{gathered}
\end{equation}
These choices give $\alpha=-0.1$, $\beta=0.4$, $x^*=0.5$, and $y^*=0.25$. Authors favor acceptance-oriented work once proxy assessment exceeds one quarter of evaluations; reviewers favor proxy assessment once acceptance-oriented authors exceed one half of the population. Cheaper production makes authors switch before proxy review becomes dominant. The feedback can therefore begin while most evaluations are still careful.

\begin{figure}[tb]\centering\includegraphics[width=\linewidth]{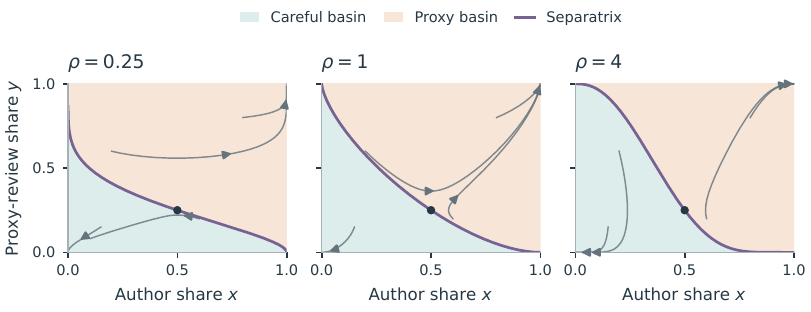}\caption{Basin membership under three reviewer adaptation speeds. Teal denotes the careful basin and orange the proxy basin. The purple curve is the analytical stable separatrix; the dark point is the saddle. Gray curves and arrows show representative trajectories. All 5,043 numerical classifications agree with the analytical boundary. The initial-condition grid excludes the absorbing edges.}\label{fig:basins}\end{figure}

Figure~\ref{fig:basins} shows how reviewer adjustment speed changes the outcome. For $\rho=0.25,1,4$, respectively, 462, 532, and 642 of the 1,681 tested starts reach the careful regime. The payoff thresholds are identical, but the order in which the populations change affects which practice becomes established.

When exploratory authors remain sufficiently common, faster movement toward careful review protects their acceptance prospects before acceptance-oriented production spreads. Once acceptance-oriented authors are sufficiently common, faster reviewer adjustment instead accelerates proxy assessment. Speed amplifies the response to existing research practices. The larger careful basin in this example reflects the balance of these two effects; the basin boundary shows which initial communities recover and which become locked into the proxy cycle.

\begin{figure}[tb]\centering\includegraphics[width=\linewidth]{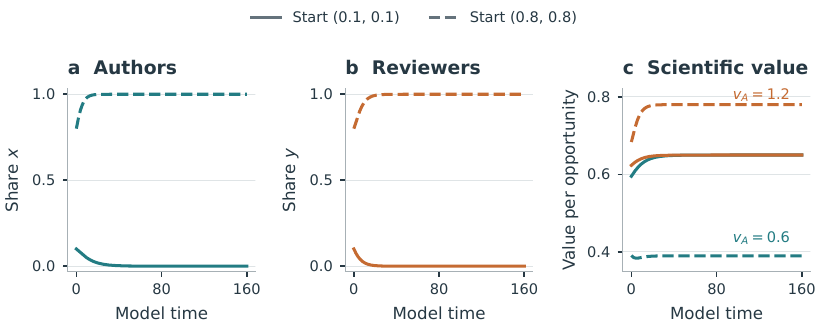}\caption{The same evolutionary behavior under different scientific-value assumptions. Solid trajectories begin near the careful corner; dashed trajectories begin near the proxy corner. Panels (a) and (b) separate author and reviewer shares. Panel (c) changes $v_A$ while holding every behavioral parameter fixed. The proxy equilibrium has lower value when $v_A=0.6$ and higher value when $v_A=1.2$. Time is in model units.}\label{fig:welfare}\end{figure}

Figure~\ref{fig:welfare} makes the scientific cost visible. With $v_A=0.6$, accepted value falls from 0.65 in the careful regime to 0.39 in the proxy regime, a 40\% loss despite identical publication frequency. Authors still obtain acceptance and the venue still fills its scientific record, but the community receives less value from the work it recognizes. Setting $v_A=1.2$ raises proxy value to 0.78 without changing behavior, confirming that contribution quality determines the damage. \textbf{In the lower-value case, publication counts give no warning and private incentives provide no correction.}

\begin{figure}[tb]\centering\includegraphics[width=\linewidth]{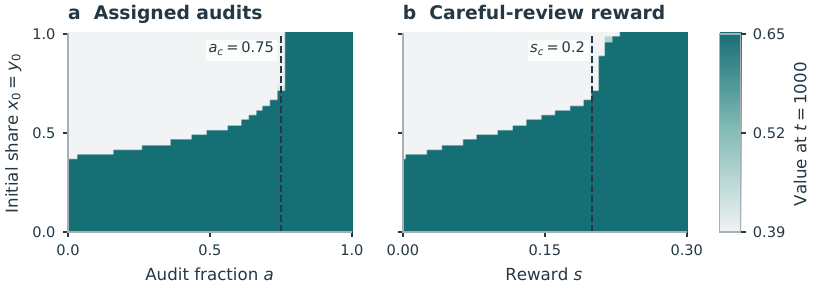}\caption{Gross scientific value at time 1,000 under permanent audits and careful-review rewards. Initial states satisfy $x=y$ and vary vertically. Dark dashed lines mark $a_c=0.75$ and $s_c=0.2$. The common value scale permits comparison within the baseline units. Policy resource costs are not subtracted, and the panels do not identify a welfare-optimal policy.}\label{fig:policies}\end{figure}

Figure~\ref{fig:policies} compares assigned careful audits and rewards for careful reviewing. Sufficient recovery thresholds are $a>0.75$ and $s>0.2$. The surfaces show gross accepted value at time 1,000 across policy levels and initial states with $x=y$; resource costs are assessed separately in Section~\ref{sec:policy}.

Weak support can protect a community near the careful regime yet leave an established proxy cycle intact. Sufficient audits make acceptance-oriented production unattractive even under widespread proxy review; sufficient rewards make careful checking worthwhile even under high workload. Both break the reinforcement, but recovery takes time. Near the thresholds, the incentive to change is small, explaining the slow response visible in the surfaces.

Temporary support succeeds when it lasts long enough to make changed practices sustain one another. With reward $s=0.3$ from $(0.8,0.8)$, the first successful duration on the one-unit test grid is $\TemporaryDuration$ model-time units. Earlier withdrawal allows ordinary incentives to restore the proxy regime. After the basin boundary is crossed, careful review and exploration continue to spread without the reward. \textbf{The policy must carry the community beyond dependence on support}, rather than merely produce improvement while funding remains available.

\begin{table}[tb]
\caption{Principal comparisons in the illustrative game. Values are per submission opportunity unless the row states a production rate. Policy thresholds are strict sufficient conditions; equality requires separate analysis.}\label{tab:mainresults}\small
\begin{tabularx}{\linewidth}{@{}p{.33\linewidth}X@{}}\toprule
Comparison & Result and interpretation\\\midrule
Baseline coordination & $x^*=0.5$, $y^*=0.25$; two stable corners\\
Careful versus proxy value & $0.65$ versus $0.39$ for $v_A=0.6$\\
Beneficial proxy comparison & Proxy value $0.78$ for $v_A=1.2$; behavior unchanged\\
Permanent support & $a>0.75$ or $s>0.2$ is sufficient for interior recovery\\
Temporary reward & First successful tested duration $43$ for $s=0.3$ from $(0.8,0.8)$\\
Higher production rate & $n_A=2$ changes incentives and removes careful-corner attraction\\\bottomrule
\end{tabularx}
\end{table}
Table~\ref{tab:mainresults} separates the forces behind these outcomes. Higher contribution quality improves scientific value directly; higher production rates change what authors find profitable; temporary support changes which practices are established when ordinary incentives resume. Reversing the harmful cycle requires changing its incentives or moving the community into a state that can sustain careful evaluation.

\subsection{Robustness and alternative specifications}
We test alternative specifications from $(0.1,0.1)$ and $(0.8,0.8)$ through time 4,000. The central question is whether research and review still reinforce one another, or whether a changed incentive interrupts that connection.

Doubling acceptance-oriented production to $n_A=2$ with $n_T=1$ removes careful-corner attraction in the tested rate specification. Producing more papers makes the strategy privately attractive even when careful review remains available. Lowering the cost of research production can therefore undermine the careful regime without any initial deterioration in reviewing. This experiment adjusts author utility and output consistently, as specified in~\eqref{eq:rateutility}.

Scarce publication slots create another pressure. With capacity $K=0.3$, the lower start approaches $x\simeq0.25$, $y\simeq0$, with accepted value 0.2775. Careful review persists, but proportional acceptance thinning reduces exploration's reward until cheaper production allows acceptance-oriented authors to coexist with it. The upper start retains the proxy regime, with value 0.18. Improving review effort alone does not remove the production incentives created by limited acceptance capacity; the edge calculation and scenario-specific units are given in Appendix~\ref{app:numerical}.

The two contrasting regimes persist under the tested nonlinear workload functions and small strategy-switching rates. Occasional switching leaves a minority using the alternative practice without dissolving the dominant feedback. Removing reviewer feedback or reversing the author response instead sends both tested starts toward careful review: the harmful cycle depends on mutual reinforcement. When the value of careful checking depends on composition in the tested alternative, trajectories keep cycling rather than settling. Thus the robust concern is the incentive connection itself, rather than a universal prediction of complete proxy adoption.

Analytical basin checks, residual diagnostics, and tighter-tolerance runs support these comparisons. The appendix reports verification details, endpoint values, and solver sensitivity; saved inputs and outputs reproduce every figure.

%% file: figures/generated_results.tex
\newcommand{\TemporaryDuration}{43.0}
\newcommand{\SolverDifference}{1.4e-07}

%% file: sections/09_discussion.tex
\section{Discussion}\label{sec:discussion}
The coordination result warns that better individual reviews may be insufficient once research production and evaluation sustain each other's weaknesses. Researchers can face pressure to abandon uncertain questions, while reviewers face pressure to reduce checking, even when both would prefer a more productive scientific community. \textbf{Reform must therefore change the rewards and costs that reproduce these practices.} The following three architectures reconsider who supplies judgments, what evidence settles a dispute, and when publication rewards are committed. Each draws on academic precedents; their integration into conference review is a proposal, rather than an outcome established by our simulations.

The common objective is to make independently valuable work and consequential checking privately sustainable. If $r\geq0$ represents a credible differential reward for validated exploratory contributions, an idealized sufficient target is
\begin{equation}\label{eq:idealgame}
\begin{aligned}
\max_{y\in[0,1]}\{\alpha+\beta(1-a)y-r\}&<0,\\
\max_{x\in[0,1]}\{\gamma+\kappa x-s\}&<0.
\end{aligned}
\end{equation}
For positive feedback, these conditions reduce to $r>\alpha+\beta(1-a)$ and $s>\gamma+\kappa$. An author would then find a validated exploratory contribution worthwhile even when proxy review is widespread, and a reviewer would find careful checking worthwhile even under high production pressure. Neither side would need to wait for the other to change first. Both payoff gaps become uniformly negative, giving interior convergence to the exploratory--careful configuration. This target is appropriate when its net scientific value is higher; it rewards validated contribution rather than unconventionality alone. The reward $r$ and the architectures below are proposed extensions, not simulated interventions.

One possibility is an \emph{auditable AI committee with an incentivized human jury}. Instead of assigning each paper several conventional reviews, a venue could deploy agents with distinct tasks: reconstructing the argument, checking technical claims, assessing prior work, and challenging the proposed decision. They would produce an evidence dossier and explicit disagreements rather than an opaque average score. Human volunteers, selected for relevant expertise and screened for conflicts, would resolve consequential disputes and inspect a random sample of apparently uncontroversial cases. Volunteers would receive bounded honoraria, review credits, or professional recognition for substantiated checks; the institution would retain human decision responsibility and an appeal route. The debate proposal of \citet{irving2018} supplies a precedent for using competing agents to expose information to a human judge, while structured reviewer recruitment \citep{stelmakh2020} supports developing the human oversight pool. Impartiality would be an institutional requirement to test, rather than a property conferred by using several agents.

Impartiality would instead be an audited property: agents would first assess anonymized work under a published rubric, and the institution would measure responses to controlled changes in affiliation, wording, and research topic against independent expert and artifact-based assessments. Several agents sharing a model or training distribution would not count as independent confirmation. Rewards would depend on independently adjudicated evidence, not agreement with the AI recommendation. Peer-prediction theory offers a formal precedent for eliciting informative effort and reports under specified assumptions \citep{miller2005}; it does not remove collusion or uninformative equilibria automatically. In the game, this architecture seeks to lower the cost of careful evaluation and increase its effective supply, while maintaining an externally financed audit fraction $a$. Its decisive test would be verified-error detection and selection quality per unit of human oversight, including failures on novel work.

A second architecture is a \emph{claim-level review and replication market}. Publication would become entry into a versioned scientific record, while certification and subsequent recognition would depend on the evidence attached to individual claims. Authors would declare consequential claims and feasible tests. Independent contributors could earn rewards for a successful reproduction, a decisive counterexample, or a documented correction, with the same evidential standard applied to confirmation and refutation. A separate forecasting mechanism could estimate which claims most need checking, and the venue would allocate a replication budget accordingly. Prediction markets have aggregated useful information about replication outcomes \citep{dreber2015}; open-research standards provide a basis for making the relevant artifacts inspectable \citep{nosek2015}. Forecast prices would guide attention rather than certify truth or determine acceptance by themselves.

The central change is to reward resolving uncertainty after publication rather than merely producing favorable pre-publication judgments. Independently settled tests could supply the effective contribution reward $r$ and a review-effort reward $s$, tightening the connection between career benefit and validated value. The forecasting and testing roles would need separation, disclosed interests, independently selected adjudicators, and rewards that do not depend on a desired outcome. Claims with no meaningful near-term settlement rule, including many conceptual contributions, would retain expert assessment rather than receive artificial probability scores. A pilot should compare correction speed, informative replications, and total verification cost, while checking whether easily testable incremental claims crowd out difficult but important questions.

A third possibility is a \emph{prospective publication contract with an exploration allocation}. For suitable empirical work, the venue would evaluate the question, identification strategy, and planned evidence before results are known, then offer a publication commitment conditional on faithful execution. AI agents could check consistency between the agreed protocol, artifacts, and final claims, while human panels retain responsibility for conceptual significance and justified deviations. Registered Reports provide the institutional precedent \citep{chambers2013,chambers2022}. A separate, explicitly budgeted exploration allocation could admit technically credible but uncertain projects without requiring a precise ranking of their expected impact. Among projects exceeding a declared quality threshold, a transparent lottery could allocate scarce mentoring, verification support, or presentation opportunities. The modified funding-lottery proposal of \citet{fang2016} offers a precedent for allocating support when precise merit rankings are unreliable.

This architecture would separate a project's right to report a well-executed investigation from whether its result is positive, fashionable, or immediately persuasive. Prospective commitment could weaken the acceptance advantage of outcome-oriented presentation, reducing $\beta$, while protected exploration and follow-up recognition could increase the effective return to exploratory work. Theory and open-ended discovery require a different contract, based on declared proof obligations or milestones rather than fixed hypotheses. The exploration allocation would remain limited, with validity thresholds and follow-up evaluation; randomization would express uncertainty among eligible projects rather than replace substantive screening. Its evaluation would compare validity, topic diversity, and later contribution under a prespecified horizon, with an explicit accounting of support costs.

These proposals target the incentives that keep the harmful cycle operating. The committee makes consequential checking easier to supply, the market rewards resolving claims, and the prospective contract protects research from pressure to produce immediately persuasive results. Each seeks to make discovery and verification worthwhile to the people carrying them out. Their component precedents motivate staged pilots measuring independent validity, recognition of exploratory work, resource use, unequal burdens, and gaming. The architectures remain proposals: binary strategies and illustrative parameters do not predict their performance at a particular venue. Their value is to direct reform toward the feedback sustaining the problem, beyond simply generating papers and reviews faster.

%% file: sections/11_conclusion.tex
\section{Conclusion}\label{sec:conclusion}
The AI research--review cycle can damage a scientific community while continuing to reward its participants. Our evolutionary game identifies how acceptance-oriented production and proxy evaluation reinforce one another, making a lower-value research regime privately stable. The community can maintain publication activity while losing scientific contribution, and isolated improvements may fail because each population sustains the other's incentives. The analytical basin boundary explains this persistence and identifies when temporary support can produce lasting recovery. Careful audits and reviewer rewards offer two mechanisms for breaking the feedback; the proposed institutional architectures extend that objective to how claims are checked and contributions recognized. The central warning is that faster production and evaluation can entrench a system that rewards passing review more than advancing knowledge. Reform should make validated discovery and consequential checking privately sustainable.

%% file: sections/10_appendix.tex
\section{Mathematical analysis}\label{app:proofs}
This appendix supplies the arguments behind the main stability and policy claims. It begins with existence and boundary behavior, then uses the first integral to establish the basin structure before deriving convergence under intervention.

\subsection{Invariant state space and equilibria}
The right-hand side of~\eqref{eq:dynamics} is polynomial, hence locally Lipschitz. Each boundary face has zero normal component. Uniqueness prevents a trajectory from crossing an invariant edge, so solutions starting in the square remain there. Compactness then supplies global forward existence. An interior solution cannot reach an edge at finite time; equivalently, the log-odds variables have bounded derivatives over finite time intervals.

For an equilibrium, each coordinate is either on its boundary or its payoff gap vanishes. If $\alpha,\alpha+\beta,\gamma,\gamma+\kappa$ are all nonzero, the only boundary equilibria are the four corners. For $\beta\kappa\neq0$, the only possible interior equilibrium is~\eqref{eq:thresholds}. It exists only if both coordinates are strictly between zero and one.

The general Jacobian is
\begin{equation}\label{eq:jacobian}
 J(x,y)=\begin{pmatrix}
 (1-2x)(\alpha+\beta y)&\beta x(1-x)\\
 \rho\kappa y(1-y)&\rho(1-2y)(\gamma+\kappa x)
 \end{pmatrix}.
\end{equation}
At a corner its off-diagonal entries vanish, giving Table~\ref{tab:eigen}. At an interior equilibrium the diagonal vanishes and the determinant is
$-\rho\beta\kappa x^*(1-x^*)y^*(1-y^*)$. A negative determinant gives one positive and one negative eigenvalue. This proves the saddle part of Proposition~\ref{prop:local}.

To establish the center claim when $\beta\kappa<0$, differentiate the first integral. At the equilibrium,
\begin{equation}
 F''(x^*)=\frac{\kappa}{x^*(1-x^*)},\qquad
 G''(y^*)=\frac{\beta}{y^*(1-y^*)}.
\end{equation}
The Hessian of $H=F-G/\rho$ is definite when the signs of $\beta$ and $\kappa$ differ. Nearby regular level sets are compact closed curves surrounding the strict extremum. The vector field is nonzero on those curves and tangent to them; the corresponding orbits are periodic. Thus the equilibrium is a nonlinear center for this specific affine system. Purely imaginary eigenvalues alone would not prove that conclusion for a general perturbation.

Endpoint equalities create additional equilibria. If $\alpha=0$, the entire bottom edge $y=0$ is stationary. If $\alpha+\beta=0$, the entire top edge is stationary. If $\gamma=0$, the left edge is stationary; if $\gamma+\kappa=0$, the right edge is stationary. Intersections of these conditions may produce several stationary edges. If $\alpha=\beta=0$, the author coordinate is stationary everywhere; reviewer equilibria are then determined conditional on that coordinate. If both payoff functions vanish identically, every point is an equilibrium. These cases are excluded from the strict coordination theorem.

\subsection{First integral and the coordination theorem}
For an interior trajectory,
\begin{align}
 \frac{\dd F}{\dd t}&=(\gamma+\kappa x)(\alpha+\beta y),\\
 \frac{1}{\rho}\frac{\dd G}{\dd t}&=(\alpha+\beta y)(\gamma+\kappa x).
\end{align}
Their difference is zero, proving~\eqref{eq:firstintegral}. Under~\eqref{eq:bistability}, both $F$ and $G$ are strictly convex on the open interval. For example,
\begin{equation}
 F''(x)=-\gamma/x^2+(\gamma+\kappa)/(1-x)^2>0.
\end{equation}
Each has a unique minimum at its threshold and diverges to positive infinity at both endpoints. The saddle level is consequently the equality of two nonnegative departures from those minima, as in~\eqref{eq:separatrix}.

Locally, its two branches have opposite slopes because the quadratic form has opposite signs. The branch in the northwest rectangle has $\dot x>0$ and $\dot y<0$; on the southeast branch $\dot x<0$ and $\dot y>0$. These branches approach the saddle and form its stable manifold. The other two branches lie in the southwest and northeast rectangles and move away from the saddle. Extending the stable branch gives a decreasing curve joining $(0,1)$ to $(1,0)$ in the closure of the square.

The southwest threshold rectangle is forward invariant: below both thresholds, both coordinates decrease, and its upper and right boundaries point into it. Any trajectory entering it has limits for both coordinates; the payoff signs force both limits to zero. The northeast rectangle is similarly invariant, and its trajectories converge to $(1,1)$. In either off-diagonal rectangle, the two coordinates move monotonically in opposite directions until the trajectory exits or approaches the saddle. It cannot remain there with any other interior limit, because one of the payoff gaps would be nonzero. It also cannot reach an outer boundary while staying in that rectangle: the signs point away from the relevant outer endpoints. Therefore every trajectory off the stable manifold eventually enters one of the two invariant rectangles. This proves the convergence and basin claims in Theorem~\ref{thm:bistable} and Proposition~\ref{prop:separatrix}.

For both cross-effects negative, the same local calculation interchanges which corners attract. Opposite signs instead give the center described above when an interior equilibrium exists. These results require no welfare assumptions.

\subsection{Policy thresholds and convergence}
With an assigned audit, the author gap is $f_a(y)=\alpha+\beta(1-a)y$. Under coordination assumptions it is increasing in $y$. Its maximum on $[0,1]$ is $\alpha+\beta(1-a)$, which is strictly negative exactly when $a>1+\alpha/\beta$. If that holds, choose $\varepsilon>0$ such that $f_a\leq-\varepsilon$. For an interior solution,
\begin{equation}
 \frac{\dd}{\dd t}\log\frac{x}{1-x}=f_a(y)\leq-\varepsilon.
\end{equation}
Hence $x\to0$. Since $\gamma<0$, the reviewer gap becomes uniformly negative after a finite time, and the analogous log-odds argument gives $y\to0$. Corner eigenvalues give the stated local stability threshold. At $y=0$, exposure is already zero and $f_a(0)=\alpha$ for every audit fraction. This proves Proposition~\ref{prop:audit}.

For a careful-review reward, the maximum reviewer gap is $\gamma+\kappa-s$. If $s>\gamma+\kappa$, then $y\to0$ by the same argument. As $\alpha<0$, the author gap is eventually negative and $x\to0$. For $s<s_c$, the proxy corner has two strictly negative eigenvalues and the reviewer nullcline shifts to $(s-\gamma)/\kappa$. This proves Proposition~\ref{prop:subsidy}.

At $a=a_c$, $f_a(1)=0$ and the top edge is stationary in its author direction. At $s=s_c$, the right-edge reviewer gap vanishes. These cases can have nonisolated equilibria or long transients. We make no blanket strict-attraction claim at equality. Also, all pure corners remain exact fixed points under the unmutated dynamics even when they become unstable; global convergence statements above concern interior initial conditions.

\subsection{Scientific value and throughput}
At $(0,0)$, equations~\eqref{eq:quantity}--\eqref{eq:welfare} give $(Q,W)=(p_T^C,p_T^Cv_T)$. At $(1,1)$ they give $(p_A^M,p_A^Mv_A)$. Comparing these expressions proves Proposition~\ref{prop:welfare}. If $v_i^j$ varies by protocol, the conditional means at the two corners are $v_T^C$ and $v_A^M$ and the same argument applies.

For throughput, subtract $n_T(Bp_T-c_T)$ from $n_A(Bp_A-c_A)$ and collect the constant and $y$ terms. This gives $\alpha_n$ and $\beta_n$ in Section~\ref{sec:welfare}. It also shows why throughput changes cannot be represented solely by multiplying the social objective. In the capacity model, the common thinning multiplier depends on the state, so that collection no longer produces a constant author intercept independent of $x$.

\section{Model extensions}\label{app:extensions}
The following supporting analysis develops the full incentive accounting, equilibrium classification, and alternative welfare specifications.
\subsection{Incentive interpretation and units}
The evidence motivates two interacting choices: how research is produced and how it is assessed. We now express those choices as private payoff differences, beginning with a consistent unit of analysis and then linking reviewer incentives to workload. This formulation makes the conditions supporting feedback explicit.

\subsubsection{Population states and the unit of analysis}
Consider a large population of authors and a large population of evaluators. An author chooses a research strategy before submitting; an evaluator chooses an assessment protocol before producing a judgment. The baseline represents one submission opportunity per author in an adjustment period. The share of authors using acceptance-oriented research is $x$, and the share of evaluations using proxy assessment is $y$. Both states lie in $[0,1]$.

An acceptance-oriented author chooses a project and presentation primarily for their expected return in the evaluation system. An exploratory author accepts greater uncertainty about recognition in pursuing a contribution. These strategies do not determine correctness or value by definition. Both can produce sound or unsound work, and either can use AI. Their scientific values will enter separately in Section~\ref{sec:welfare}.

Careful evaluation includes the effort needed to understand an argument and check its consequential claims. Proxy evaluation relies more heavily on inexpensive correlates of merit. Familiarity, presentation, standard experimental structure, or an automated assessment can serve as such correlates. Their usefulness depends on the setting. The distinction is the basis for the judgment, rather than the presence of generated text in a review.

An interaction has an expected acceptance outcome, a private reward, and a scientific-value outcome. Participants adapt after observing relative private returns. We use deterministic population dynamics as an approximation to this adaptation. Authors and reviewers may be the same people in practice; the two populations represent roles, and the baseline omits correlations between a person's behavior across those roles.

\subsubsection{Author rewards}
Let $p_i^j\in[0,1]$ be the probability that research of strategy $i\in\{A,T\}$ is accepted under evaluation $j\in\{C,M\}$. These probabilities summarize an institutional protocol, including its decision rule. The baseline has no binding acceptance quota. We treat $B>0$ as the expected private benefit of acceptance and $c_i\geq0$ as research cost per submission opportunity. Thus
\begin{equation}
 U_i(y)=B p_i(y)-c_i,\qquad p_i(y)=(1-y)p_i^C+yp_i^M.
\end{equation}
The linear mixture assumes that an opportunity encounters a protocol drawn from the current evaluation population. It does not claim that a conference averages reviewer scores in this way. A multi-reviewer committee rule would generally produce a nonlinear acceptance function, for which the local incentive interpretation can still be examined.

Write $\Delta_C=p_A^C-p_T^C$ and $\Delta_M=p_A^M-p_T^M$. The relative payoff is
\begin{align}
 D_A(y)&=U_A(y)-U_T(y)=\alpha+\beta y,\\
 \alpha&=c_T-c_A+B\Delta_C,\qquad
 \beta=B(\Delta_M-\Delta_C).
\end{align}
The intercept $\alpha$ is the invasion payoff for acceptance-oriented research in a careful-review environment. The coefficient $\beta$ is a difference in relative acceptance rewards across protocols. A positive $\beta$ means that proxy reviewing improves the relative position of $A$; it does not require proxy reviewers to prefer $A$ in absolute terms. For example, both $\Delta_C$ and $\Delta_M$ could be negative while $\beta$ remains positive.

Research cost savings can also make $A$ attractive independently of any review interaction. If $c_T-c_A$ is large enough, $D_A(y)$ is positive throughout the interval even when careful reviewers favor exploratory contributions in acceptance probability. The model therefore separates a production-cost advantage from an evaluation advantage. Conflating them would make it difficult to infer which intervention could change behavior.

\subsubsection{Reviewer costs and workload}
Let $L(x)=L_0+\ell x$ denote an evaluator's load over an adjustment period. Load is measured in review-task units and must be nonnegative over the state space. It can change with submission volume or the verification demands of submissions. At the institutional scale, production rates can generate $\ell>0$; at the assignment scale, a fixed cap or reduced verification burden could instead make $\ell$ zero or negative.

Suppose proxy assessment saves $\delta t\geq0$ payoff units per load unit. Careful assessment supplies an accuracy advantage $\Delta q$, valued privately at $\eta\geq0$, while proxy evaluation incurs an additional cost $d$. A shared reward for completing the assignment cancels from the difference. Then
\begin{equation}
 D_R(x)=U_M(x)-U_C(x)
 =\delta t(L_0+\ell x)-\eta\Delta q-d=\gamma+\kappa x,
\end{equation}
where $\gamma=\delta tL_0-\eta\Delta q-d$ and $\kappa=\delta t\ell$.

The fixed accuracy term is a simplification: it values the quality of a review protocol over the period rather than accumulating a separate benefit per checked claim. If the accuracy benefit also scales with load, its marginal contribution must be included in $\kappa$. Likewise, if the papers produced under $A$ are easier to check, load need not increase with $x$. These possibilities are not hidden in the derivation; they change a coefficient whose sign is central to the result.

Author costs are per submission opportunity, whereas reviewer costs summarize a review period. The relative adjustment factor below accounts for the independent scales of these two payoff differences. A detailed accounting of production per researcher-period is introduced only in the throughput extension, where both author utility and scientific output are changed together.

\subsubsection{Adaptation}
Let strategies with higher private payoff increase their activity shares according to
\begin{equation}
 \dot x=x(1-x)(\alpha+\beta y),\qquad
 \dot y=\rho y(1-y)(\gamma+\kappa x),\qquad \rho>0.
\end{equation}
The factors $x(1-x)$ and $y(1-y)$ express the presence of both alternatives in a population. The equations preserve absent strategies: a population at exactly $x=0$ cannot generate $A$ without switching or entry. Our convergence claims therefore specify interior initial conditions. Section~\ref{sec:numerics} also studies small symmetric switching rates.

The adaptation speed $\rho$ combines how quickly reviewers revise their behavior with the relative scale of reviewer payoffs. Model time has no empirical conversion to months or conference cycles. Rescaling all payoffs changes the speed of adjustment, not the payoff signs defining the equilibria.

\begin{table}[tb]\caption{Principal quantities and units.}\small
\begin{tabularx}{\linewidth}{@{}p{.19\linewidth}X@{}}\toprule
Symbol & Interpretation\\\midrule
$x,y$ & Strategy shares in the author and evaluation roles\\
$p_i^j$ & Acceptance probability per submission under protocol $j$\\
$B,c_i$ & Acceptance benefit and research cost per submission opportunity\\
$\alpha,\beta$ & Author payoff intercept and evaluation interaction\\
$\gamma,\kappa$ & Reviewer payoff intercept and composition feedback\\
$\rho$ & Relative reviewer adjustment speed\\
$v_i$ & Scientific value conditional on acceptance\\
$a,s$ & Assigned careful-audit fraction and careful-review reward\\\bottomrule
\end{tabularx}\end{table}

\subsection{Equilibrium classification and basin geometry}
The payoff equations determine where each strategy can spread, but local advantages alone do not determine the population outcome. We first classify equilibria and then derive the boundary separating their basins. This distinction is essential for understanding how a cost change or temporary intervention can redirect the same community.

\subsubsection{Local behavior}
The four corners of the unit square are equilibria because a pure population contains no alternative strategy that can increase through imitation alone. When $\beta\kappa\neq0$, an interior equilibrium exists precisely when
\begin{equation}
 x^*=-\gamma/\kappa\in(0,1),\qquad
 y^*=-\alpha/\beta\in(0,1).
\end{equation}
Each threshold is a best-response switch: authors favor $A$ above $y^*$ when $\beta>0$, and reviewers favor $M$ above $x^*$ when $\kappa>0$. The crossing of these nullclines does not itself attract a population.

\begin{proposition}[Invariant state space and local classification]\label{prop:local}
Equation~\eqref{eq:dynamics} has a unique solution for each initial state in the square, and the square and its edges are forward invariant. The eigenvalues at its corners are given in Table~\ref{tab:eigen}. At an interior equilibrium, the eigenvalues satisfy
\begin{equation}
 \lambda^2=\rho\beta\kappa x^*(1-x^*)y^*(1-y^*).
\end{equation}
An interior equilibrium is a saddle when $\beta\kappa>0$. When $\beta\kappa<0$, it is a nonlinear center for this affine baseline.
\end{proposition}

\begin{table}[tb]\caption{Corner eigenvalues. Both eigenvalues must be strictly negative for local asymptotic stability; zero values require separate analysis.}\label{tab:eigen}
\centering\begin{tabular}{lll}\toprule
Corner & Author direction & Reviewer direction\\\midrule
$(0,0)$ & $\alpha$ & $\rho\gamma$\\
$(1,0)$ & $-\alpha$ & $\rho(\gamma+\kappa)$\\
$(0,1)$ & $\alpha+\beta$ & $-\rho\gamma$\\
$(1,1)$ & $-(\alpha+\beta)$ & $-\rho(\gamma+\kappa)$\\\bottomrule
\end{tabular}\end{table}

The distinction between a saddle and a center matters for interpretation. Positive cross-effects support reinforcing movement. Opposite cross-effects can generate persistent oscillation around the interior state in the unmodified model, rather than convergence. A small damping mechanism or a different update rule could change the latter behavior. We therefore do not infer stable long-run shares from imaginary eigenvalues alone; the center claim follows from the first integral below.

\begin{theorem}[Coordination regime]
Suppose $\beta>0$, $\kappa>0$, and
\begin{equation}
 \alpha<0<\alpha+\beta,\qquad
 \gamma<0<\gamma+\kappa.
\end{equation}
Then $(0,0)$ and $(1,1)$ are locally asymptotically stable, the other corners are repelling, and $(x^*,y^*)$ is a saddle. Its stable manifold divides the open square into two basins. Every interior initial state outside this manifold converges to one of the two attracting corners.
\end{theorem}

This result identifies coexistence of institutional possibilities, not coexistence of strategies at a stable interior mixture. A community with many exploratory projects and careful reviews can sustain that combination. A community organized around acceptance signals and proxy evaluation can also sustain itself under the same parameters. An isolated participant's incentive to change depends on the practices surrounding that participant.

\begin{figure}[tb]\centering\includegraphics[width=\linewidth]{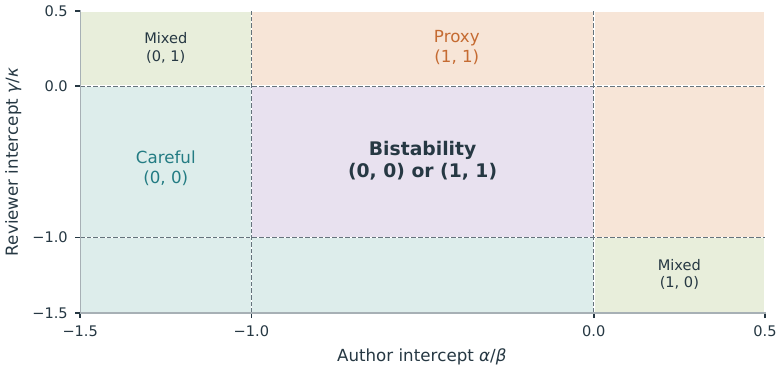}\caption{Local attracting regimes for positive cross-effects. Axes normalize the intercepts by $\beta$ and $\kappa$. The central square has two attracting corners. ``Mixed corner'' denotes opposite pure strategies in the two roles, not a stable interior mixture. Lines at zero and minus one are nonhyperbolic boundaries. The regime map is analytical, not an empirical estimate.}\label{fig:regimes}\end{figure}

The signs in~\eqref{eq:bistability} are stronger than $\beta,\kappa>0$. Positive feedback alone does not imply two equilibria. If acceptance-oriented research has a private advantage even under exclusively careful review, the first inequality fails. If careful review remains privately preferable even at the highest load, the last inequality fails. Figure~\ref{fig:regimes} displays the resulting regimes for positive cross-effects.

\subsubsection{The basin boundary}
The affine model allows a more precise statement than a phase portrait alone. Define, on the open square,
\begin{equation}
 F(x)=\gamma\log x-(\gamma+\kappa)\log(1-x),\quad
 G(y)=\alpha\log y-(\alpha+\beta)\log(1-y).
\end{equation}
Their derivatives are $F'(x)=(\gamma+\kappa x)/[x(1-x)]$ and $G'(y)=(\alpha+\beta y)/[y(1-y)]$. Hence
\begin{equation}
 H(x,y)=F(x)-\rho^{-1}G(y)
\end{equation}
is constant along every interior trajectory.

\begin{proposition}[Separatrix characterization]\label{prop:separatrix}
Under~\eqref{eq:bistability}, the stable manifold is the decreasing branch through $(x^*,y^*)$ of
\begin{equation}
 G(y)-G(y^*)=\rho\{F(x)-F(x^*)\}.
\end{equation}
It lies in the northwest and southeast threshold rectangles: $x<x^*,y>y^*$ or $x>x^*,y<y^*$. The increasing branch of the same level set is the unstable manifold and must not be used as the basin boundary.
\end{proposition}

The branch distinction prevents a common error in plotting a conserved quantity. The saddle level contains both trajectories that approach the saddle and trajectories that leave it. Only the former separate the attracting outcomes. In the lower-left threshold rectangle, both populations move toward the careful corner; in the upper-right rectangle, both move toward the proxy corner. The two off-diagonal rectangles contain the competition between these tendencies.

Changing $\rho$ leaves the nullclines and local stability signs unchanged, but it changes equation~\eqref{eq:separatrix}. Faster reviewer adjustment therefore changes which initial states reach each corner. It need not expand the same basin everywhere: a rapid response can help careful review spread when $x<x^*$ and accelerate proxy adoption when $x>x^*$. Numerical basin counts will make this dependence visible without converting a grid fraction into an empirical probability.

\subsubsection{Costs and structural change}
Let $d_A=c_T-c_A$ denote the author cost difference, holding acceptance probabilities and $B$ fixed. Within the coordination regime,
\begin{equation}\label{eq:comparative}
 \frac{\partial y^*}{\partial d_A}=-\frac1\beta,
 \qquad \frac{\partial x^*}{\partial\gamma}=-\frac1\kappa.
\end{equation}
Cheaper acceptance-oriented production lowers the proxy-review share needed for $A$ to be privately attractive. A reviewer cost change that raises $\gamma$ lowers the author share needed to make $M$ attractive. These are threshold comparisons. They do not establish a universal monotonic basin-area theorem when several coefficients change simultaneously.

The careful corner loses strict local stability at $\alpha=0$ or $\gamma=0$. The proxy corner loses it at $\alpha+\beta=0$ or $\gamma+\kappa=0$. These are structural boundaries in the game. A trajectory crossing the separatrix at fixed parameters is a different event: it changes the eventual outcome without changing the equilibrium set. We use ``path dependence'' for the coexistence of basins and reserve a bifurcation claim for a change in the vector field at a stability boundary.

\subsubsection{Degenerate and reversed feedback}
If $\beta=0$ and $\alpha\neq0$, author behavior changes monotonically according to a logistic equation, independently of $y$. If $\alpha=\beta=0$, every author share is stationary and reviewer behavior adjusts conditional on that fixed share. The reviewer statements are analogous when $\kappa=0$. An endpoint equality, such as $\alpha=0$, can create an equilibrium edge; the strict corner criteria must not be extrapolated through that equality.

When both cross-effects are negative and an interior equilibrium exists, it is again a saddle, but the attracting corners are $(1,0)$ and $(0,1)$. When their signs differ, the interior first integral has a definite Hessian and nearby level curves are closed. These alternatives show that convergence to the all-proxy corner is not a generic implication of a two-population model. It follows from particular signs and sufficiently strong relative rewards.

\subsection{Value, throughput, and capacity}
The preceding analysis identifies stable behavior. Assessing its scientific consequences requires an objective beyond acceptance rewards. We therefore introduce contribution value, distinguish it from accepted quantity, and examine how production rates and limited capacity change both incentives and output.

\subsubsection{Accepted quantity and accepted value}
Evolutionary stability concerns private rewards. It does not rank the contributions selected by a research institution. Introduce $v_i\geq0$ as expected scientific value conditional on acceptance for author strategy $i$. The scalar is a declared aggregation of validity, usefulness, and contribution over a chosen time horizon. Its interpretation must be held fixed within a comparison. It is not a review score, publication count, or synonym for novelty.

For the baseline unit of one submission opportunity, define
\begin{align}
 Q(x,y)&=xp_A(y)+(1-x)p_T(y),\\
 W(x,y)&=xp_A(y)v_A+(1-x)p_T(y)v_T.
\end{align}
When $Q>0$, $\overline V=W/Q$ is the average value of accepted work. $W$ measures the value entering the accepted record per opportunity. It does not measure all discoveries, including rejected work later disseminated elsewhere. Its use as an institutional objective consequently assumes that acceptance affects attention, resources, or incorporation into the scientific record.

The baseline assigns one conditional mean $v_i$ to each strategy across protocols. This is useful for isolating the incentive mechanism, but a careful reviewer might select a different subset of $A$ papers than a proxy reviewer. Allowing $v_i^j$ gives the more general expression
\begin{equation}
 W=x[(1-y)p_A^Cv_A^C+yp_A^Mv_A^M]
 +(1-x)[(1-y)p_T^Cv_T^C+yp_T^Mv_T^M].
\end{equation}
The dynamical analysis is unchanged when these values do not enter private payoffs. The welfare comparison can change substantially, because equal acceptance probabilities can conceal different selection quality.

\begin{proposition}[Welfare ordering of the coordination equilibria]\label{prop:welfare}
For the baseline scientific-value function, the proxy corner has lower accepted scientific value per opportunity than the careful corner if and only if
\begin{equation}
 p_A^M v_A<p_T^C v_T.
\end{equation}
Under protocol-dependent selection values, replace $v_A,v_T$ by $v_A^M,v_T^C$. Stability conditions alone impose neither ordering.
\end{proposition}

The proposition follows by evaluating $W$ at the corners. Its simplicity is useful: it identifies the extra empirical comparison required before a stable population configuration can be described as scientifically inferior. A preference interaction $\beta>0$ is not enough. Even the full coordination inequalities leave both $v_A$ and $v_T$ unrestricted.

In the numerical example, both attracting corners accept 0.65 papers per opportunity. With $v_A=0.6$ and $v_T=1$, their accepted values are 0.39 and 0.65. If $v_A$ is increased to 1.2, the same proxy corner instead produces 0.78. The population trajectories do not change, because neither value enters the payoff equations. Figure~\ref{fig:welfare} uses this comparison to show why adoption and welfare must be plotted together.

\subsubsection{Private incentives and the social objective}
An institution might compare $W$ with review resources, rather than maximize gross accepted value alone. Let $C_C(x)$ and $C_M(x)$ be social evaluation costs and $\lambda\geq0$ their conversion into value units. A possible net criterion is
\begin{equation}\label{eq:netwelfare}
 J=W-\lambda[(1-y)C_C(x)+yC_M(x)].
\end{equation}
This objective is distinct from reviewer utility. A reward for reviewing can be a transfer between participants, whereas time spent checking a claim consumes real resources. A private time cost and a social accuracy benefit need not have the same weight. The paper therefore reports behavioral thresholds without declaring an optimal subsidy or audit rate in the absence of those weights.

The gap between private and social objectives is the institutional issue. At an attracting corner, no small invading strategy can earn a larger private return. That condition can coexist with lower $W$ or lower $J$. It can also coexist with higher welfare. The welfare ranking is an additional property of the environment, not part of the definition of an equilibrium.

\subsubsection{Production rates and consistent units}
Suppose authors using strategy $i$ generate $n_i>0$ submission opportunities per researcher-period. Research costs $c_i$ remain costs per submission. Then private utility and accepted output per researcher-period become
\begin{align}
 U_i^{\mathrm{rate}}(y)&=n_i[Bp_i(y)-c_i],\label{eq:rateutility}\\
 Q^{\mathrm{rate}}(x,y)&=xn_Ap_A(y)+(1-x)n_Tp_T(y),\\
 W^{\mathrm{rate}}(x,y)&=xn_Ap_A(y)v_A+(1-x)n_Tp_T(y)v_T.\label{eq:ratewelfare}
\end{align}
The author intercept and interaction must also change:
\begin{align}
 \alpha_n&=n_A(Bp_A^C-c_A)-n_T(Bp_T^C-c_T),\\
 \beta_n&=B[n_A(p_A^M-p_A^C)-n_T(p_T^M-p_T^C)].
\end{align}
The corner welfare condition is now $n_Ap_A^Mv_A<n_Tp_T^Cv_T$. Increasing throughput can change both behavior and welfare; it is inconsistent to change only the latter and present the resulting trajectory as a prediction of rate-dependent incentives.

The rate sensitivity experiment holds the reduced reviewer coefficients fixed to isolate this direct author effect. A fully linked institutional version would also derive workload from total submissions, for example $L=L_0+\xi[xn_A+(1-x)n_T]$ for fixed reviewer supply. This would change $\gamma$ and $\kappa$ together. The simulation makes no claim that the isolated rate perturbation holds every institutional relationship constant in reality.

\subsubsection{Capacity-constrained selection}
A fixed number of acceptance slots couples authors through competition. As one transparent sensitivity case, let $K>0$ be the accepted-output cap per researcher-period. Form the unconstrained demand
\begin{equation}
 D=xn_Ap_A(y)+(1-x)n_Tp_T(y),\qquad
 r=\min\{1,K/D\},
\end{equation}
with $r=1$ at $D=0$, and replace every acceptance probability by $\widetilde p_i=rp_i$. This is proportional thinning, not a model of ranking papers by merit. It ensures accepted quantity does not exceed $K$ while preserving relative acceptance weights conditional on the state.

Author incentives now depend on both $x$ and $y$ through $r$. The affine author gap and its closed-form separatrix no longer apply. Costs are still paid on submitted work, so a capacity cap can reduce the acceptance return relative to research cost. If all papers have positive expected value, accepted value also becomes constrained by the available slots. These effects are examined numerically rather than inferred from the baseline theorem.

\subsection{Policy accounting and boundary cases}
A welfare comparison identifies which outcomes an institution might prefer; it does not show how to reach them. We now consider two levers acting on different parts of the feedback: assigned careful evaluation changes author exposure, while a review reward changes evaluator incentives. Their effects on stability are derived before considering temporary use and resource costs.

\subsubsection{Assigned careful evaluation}
Consider an institution that assigns a fraction $a\in[0,1]$ of submissions to careful assessment, independently of the voluntary protocol that would otherwise be used. The assigned assessment determines the outcome for those opportunities; it is not an extra advisory review whose recommendation can be ignored. The remaining fraction encounters the current reviewer population. Effective proxy exposure is therefore
\begin{equation}
 z=(1-a)y,\qquad D_A^a(y)=\alpha+\beta(1-a)y.
\end{equation}
The state $y$ continues to describe voluntary reviewer behavior. Consequently, $y=1$ with $a>0$ does not mean that every actual assessment is a proxy assessment.

We hold voluntary reviewer payoffs fixed in this first policy comparison. The institution supplies audit capacity externally; it does not remove that capacity from the same reviewers and then assume their costs are unchanged. If assigned audits crowd out voluntary checking, the reviewer coefficients require an additional adjustment. This distinction separates the behavioral benefit of an audit from its resource feasibility.

\begin{proposition}[Audit threshold]\label{prop:audit}
Under the coordination conditions of Theorem~\ref{thm:bistable}, define
\begin{equation}
 a_c=1+\alpha/\beta\in(0,1).
\end{equation}
The proxy corner remains strictly locally attracting when $a<a_c$ and loses strict stability at $a=a_c$. If $a>a_c$, every interior trajectory converges to $(0,0)$. The author invasion payoff at the careful corner remains $D_A^a(0)=\alpha$, independently of $a$.
\end{proposition}

An audit changes how rewarding proxy-compatible work can become when proxy assessment is prevalent. It cannot improve the author invasion payoff in an environment that already uses only careful evaluation. When $a<a_c$, the author threshold rises to $-\alpha/[\beta(1-a)]$ while the reviewer threshold is unchanged. This weakens the reinforcing loop without necessarily eliminating the proxy basin.

The result gives a sufficient permanent policy under the baseline assumptions. It does not say that auditing just above $a_c$ produces rapid recovery. As the payoff margin approaches zero, movement away from the proxy corner can be slow. A finite simulation horizon can therefore show substantial proxy adoption even when the asymptotic theorem predicts eventual convergence to the careful corner.

\subsubsection{A reward for careful reviewing}
Let a careful reviewer receive an additional payoff $s\geq0$ over the review period. Then
\begin{equation}
 D_R^s(x)=\gamma+\kappa x-s.
\end{equation}
Unlike the audit, this policy changes reviewers' willingness to supply careful work. It does not directly change the acceptance probabilities associated with either protocol.

\begin{proposition}[Careful-review incentive threshold]\label{prop:subsidy}
Under the coordination conditions, define $s_c=\gamma+\kappa>0$. If $s>s_c$, every interior trajectory converges to $(0,0)$. If $0\leq s<s_c$, the proxy corner remains locally attracting and the reviewer threshold is $x_s^*=(s-\gamma)/\kappa$. At equality the proxy corner is nonhyperbolic and the strict stability classification is insufficient.
\end{proposition}

The sufficient incentive exceeds the largest private advantage of proxy evaluation over the state space. This is a demanding policy: it makes careful reviewing privately preferable even at $x=1$. Smaller rewards can still matter by shifting a basin boundary. The institution may thus obtain a lasting behavioral effect without paying a permanent dominance-inducing reward, but the effectiveness of a temporary policy depends on the initial state and duration.

\subsubsection{Temporary intervention and release}
Suppose a policy is applied on $[0,T]$ and then removed. Let $\Phi_T^P(x_0,y_0)$ be the state reached under that policy. Subsequent behavior follows the baseline game from this release state. If the release state lies in the careful basin, the population approaches the careful corner after withdrawal. If it lies in the proxy basin, proxy-oriented behavior returns. A release state exactly on the separatrix approaches the saddle in the deterministic model and is sensitive to perturbations.

This is a state-dependent intervention criterion rather than a new equilibrium theorem. The experiment in Section~\ref{sec:numerics} applies a reward of $s=0.3$ to a population initially at $(0.8,0.8)$, varies its duration, and tests the released trajectory. Model duration is not calendar time. The conserved baseline quantity can locate the separatrix, but identifying the stable branch remains essential.

\subsubsection{Costs and combined policies}
With an audit, gross scientific value is evaluated at exposure $z=(1-a)y$. The share of opportunities receiving careful assessment is $a+(1-a)(1-y)$, and the proxy share is $(1-a)y$. If unit social costs are $k_C$ and $k_M$, a net criterion is
\begin{equation}\label{eq:auditnet}
 J_a=W(x,z)-\lambda\{[a+(1-a)(1-y)]k_C+(1-a)yk_M\}.
\end{equation}
Additional assignment or audit infrastructure costs can be added to this expression. For the reward policy, expenditure per normalized review period is $s(1-y)$; this is a budget cost, whereas its social cost depends on financing and opportunity costs. The code reports gross scientific value and behavioral outcomes without inventing a conversion factor $\lambda$.

When both policies operate, the proxy-corner eigenvalues become
\begin{equation}
 -[\alpha+\beta(1-a)],\qquad -\rho[\gamma+\kappa-s].
\end{equation}
Either a sufficiently strong audit or a sufficiently strong reward removes its strict attraction. Subthreshold combinations can nevertheless move a given population across the modified basin boundary. The global benefit must still be assessed using scientific value and resource costs. If the proxy configuration produces greater scientific value, moving away from it can reduce the gross objective.

\subsection{Nonlinear response and switching}
The baseline proofs depend on affine, separable payoff gaps. This appendix explains what changes under alternative reviewer responses and strategy switching, and records the numerical accounting needed to interpret those comparisons.

\subsubsection{Nonlinear reviewer response}
For a differentiable function $h$, replace the reviewer gap by $g(x)=\gamma+\kappa h(x)$. When $g$ has a unique interior zero and $g'(x^*)>0$, the interior Jacobian has the same off-diagonal form with $\kappa$ replaced by $g'(x^*)$. The corner signs are determined by $g(0)$ and $g(1)$. An increasing response can thus preserve local coordination while changing its global geometry. With multiple roots, there can be additional equilibria and the baseline theorem does not apply without further analysis.

Separability also preserves a first integral of the form
\begin{equation}
 H_h(x,y)=\int^x\frac{g(u)}{u(1-u)}\,\dd u
 -\frac1\rho\int^y\frac{\alpha+\beta v}{v(1-v)}\,\dd v.
\end{equation}
The numerical implementation integrates the vector field directly rather than evaluating these integrals. The saturating and convex examples are locally matched at $x=1/2$, but their endpoint gaps differ. This makes them a test of the global workload assumption rather than a pure rescaling of speed.

The quality-benefit perturbation subtracts $q_1(x-1/2)$ from $g(x)$. For the linear case, it changes the slope from $\kappa$ to $\kappa-q_1$ and the intercept from $\gamma$ to $\gamma+q_1/2$. At $q_1=0.6$ in the baseline, the slope is $-0.2$ and the interior reviewer threshold remains $x=1/2$. The interior equilibrium is then a center. Endpoints from this scenario can retain a nonzero residual and must not be described as convergence.

\subsubsection{Switching and inward motion}
For $\mu>0$ in~\eqref{eq:mutation}, the author normal derivative is $\mu$ at $x=0$ and $-\mu$ at $x=1$. The reviewer normal derivatives are $\rho\mu$ and $-\rho\mu$. Hence the square remains forward invariant and no corner is an equilibrium. A sufficiently small perturbation preserves each hyperbolic equilibrium locally by the implicit function theorem; the attracting corners move into the square. That local observation does not determine a universal switching rate at which bistability disappears.

In particular, one should not classify switching endpoints by their distance to baseline corners. The robustness CSV instead preserves the numerical endpoint and residual for each stated rate and initial condition. The plotted values are finite-time outcomes. No stochastic uncertainty bands are shown because the experiment has no random sampling.

\section{Additional numerical results}\label{app:numerical}\label{app:costs}
This appendix retains full specifications, endpoint comparisons, and supplementary sensitivity figures.
The analytical results identify equilibrium and policy conditions. The numerical study makes their geometry and finite-time consequences visible, then tests which patterns survive changes in workload, switching, production, and capacity. Each comparison retains its declared units and separates a simulated endpoint from an asymptotic claim.

\subsection{Design and baseline}
The baseline uses normalized illustrative inputs to examine when cheaper acceptance-oriented production and proxy evaluation sustain each other. We set
\begin{equation}
\begin{gathered}
 B=1,\quad c_A=0.1,\quad c_T=0.2,\\
 p_A^C=0.45,\quad p_T^C=0.65,\quad
 p_A^M=0.65,\quad p_T^M=0.45,\\
 \gamma=-0.2,\quad\kappa=0.4,\quad\rho=1,
 \quad v_A=0.6,\quad v_T=1.
\end{gathered}
\end{equation}
Thus $\alpha=-0.1$, $\beta=0.4$, $x^*=0.5$, and $y^*=0.25$. These values give a simple coordination example with substantial basins for both attracting corners. The author threshold is lower than the reviewer threshold because the research-cost difference partially offsets the careful protocol's acceptance preference.

The basin experiment uses a $41\times41$ equally spaced grid on $[0.01,0.99]^2$, excluding absorbing boundaries. Each initial state is integrated to time 1,000, with relative tolerance $10^{-8}$ and absolute tolerance $10^{-10}$. A trajectory is assigned to an equilibrium only if both its maximum coordinate distance and its vector-field residual are less than $10^{-5}$. Unresolved grid points are reintegrated from their original states to time 4,000 at tolerances $10^{-10}$ and $10^{-12}$. The output retains any remaining unresolved states rather than assigning them to the nearest corner.

The experiment is deterministic. The configuration includes a seed for reproducibility of future stochastic variants, but no random draws are used in the reported results. Counts over the initial-condition grid summarize geometric basins under an arbitrary uniform grid; they are not probabilities that a conference will follow a particular path.

\subsection{Basin geometry and adjustment speed}
Figure~\ref{fig:basins} plots basin membership for reviewer adaptation speeds $\rho=0.25,1,4$. All three share the same saddle and the same corner stability signs. Their basin boundaries differ because the relative speed changes the balance of movement in the off-diagonal threshold rectangles. Table~\ref{tab:basins} reports the resulting counts from generated numerical output.

\begin{table}[tb]\caption{Basin counts out of 1,681 initial conditions for each reviewer adaptation speed. These are grid counts, not estimated population frequencies.}\label{tab:basins}\centering
\input{figures/basin_table}
\end{table}

Faster reviewer adjustment amplifies the response to the current author population: it favors proxy assessment above $x^*$ and careful assessment below it. The tested grid yields a larger careful basin at higher $\rho$, while equation~\eqref{eq:separatrix} identifies which starts are redirected. The institutional issue is who adjusts first, not speed alone.

\subsection{Value and intervention effects}
Figure~\ref{fig:welfare} holds all behavioral parameters fixed and varies $v_A$. When $v_A=0.6$, the careful and proxy corners deliver accepted value 0.65 and 0.39 per opportunity. When $v_A=1.2$, proxy value is 0.78. Equal accepted quantity at the two corners makes the distinction particularly clear: the value assumptions, rather than a change in publication count, reverse the ranking.

For the baseline, the sufficient permanent policy thresholds are $a_c=0.75$ and $s_c=0.2$. Figure~\ref{fig:policies} sweeps audit coverage from zero to one and rewards from zero to 0.3, using 41 levels of each. At each policy level, 41 initial states on the diagonal $x=y$ are integrated to time 1,000. The panels report gross scientific value at that finite horizon. They do not optimize a net resource objective.

The response surfaces show both a dependence on the initial state below the sufficient thresholds and a weakening of the proxy regime near those thresholds. Close to a zero eigenvalue, finite-time outcomes can differ from the limiting corner prediction. The dashed lines therefore mark analytical stability boundaries, not promises of recovery within the plotted horizon. Capacity for the assigned careful assessments is an external resource in this experiment.

For a temporary reward of $s=0.3$ starting at $(0.8,0.8)$, we evaluate durations from zero to 100 in one-unit increments. The first tested duration whose released trajectory converges to the careful corner is $\TemporaryDuration$ model-time units. This locates the switch only to the grid resolution; it is not an estimate of the exact critical duration. The release coordinates and final states are retained in the reproducibility files, allowing the state-transfer criterion to be checked directly.

\subsection{Alternative specifications}
The robustness analysis begins from $(0.1,0.1)$ and $(0.8,0.8)$ and integrates to time 4,000. It reports final shares, scientific value, and residuals for each specification. These endpoints are observations about simulated paths, not automatic declarations that every modified system has the baseline equilibrium structure.

Two workload functions replace $x$ in the reviewer gap:
\begin{equation}\label{eq:nonlinear}
 h_s(x)=\tfrac12+\tfrac12\tanh[2(x-\tfrac12)],\qquad
 h_c(x)=x+(x-\tfrac12)^2.
\end{equation}
Both match the linear specification's value and slope at $x=1/2$. The first saturates and the second is convex. Matching locally helps separate a change in global shape from an arbitrary shift in the interior threshold. For either function, local corner stability depends on the endpoint reviewer gaps, while the interior saddle calculation uses $\kappa h'(x^*)$.

To represent changing accuracy benefits, a composition term $-q_1(x-1/2)$ is added to the reviewer gap, with $q_1=0.6$. This makes the net cross-effect negative in the stated example. It deliberately breaks the reinforcing-feedback assumption rather than providing another version of it. Likewise, the reversed-author case sets $p_A^M=0.25$ and $p_T^M=0.85$, making $\beta<0$. Zero reviewer feedback is tested by setting $\kappa=0$ while retaining $\gamma=-0.2$.

Symmetric switching is introduced through
\begin{equation}\label{eq:mutation}
\begin{aligned}
 \dot x&=x(1-x)D_A(y)+\mu(1-2x),\\
 \dot y&=\rho\{y(1-y)D_R(x)+\mu(1-2y)\},
\end{aligned}
\end{equation}
for $\mu=10^{-4},10^{-3},10^{-2}$. Switching makes the vector field point inward at the boundaries. Small rates move the attracting states away from exact corners and permit absent strategies to reappear. Their precise stability properties require the modified Jacobian; the baseline invariant is no longer conserved.

\begin{figure}[tb]\centering\includegraphics[width=\linewidth]{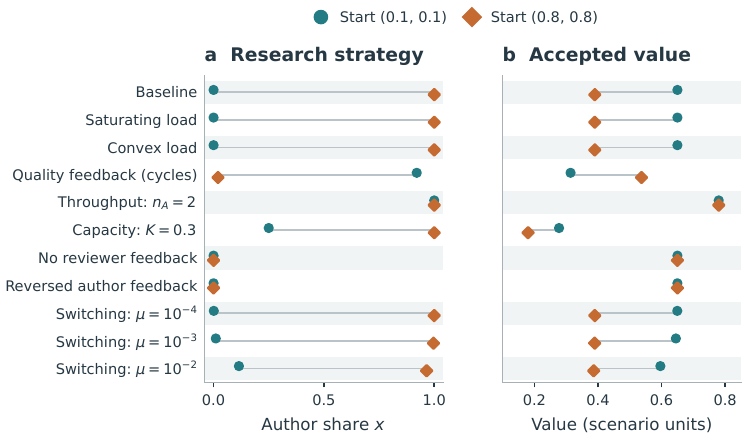}\caption{Finite-time endpoints for alternative specifications from two initial states. Left: acceptance-oriented author share. Right: accepted scientific value in each scenario's declared units. Results at time 4,000 are retained with their vector-field residuals in the CSV output. A finite endpoint, especially under opposing feedback, should not be interpreted as a stable equilibrium without the corresponding residual and stability analysis.}\label{fig:robustness}\end{figure}

The rate experiment sets $n_A=2,n_T=1$ and changes author utility as in~\eqref{eq:rateutility}. The capacity experiment uses $K=0.3$ with baseline production rates. As explained in Section~\ref{sec:welfare}, proportional thinning changes the author gap through the current state. The scientific-value units in a rate scenario are per researcher-period; the baseline is per submission opportunity. Figure~\ref{fig:robustness} labels the reported values as scenario-specific for this reason.

These changes produce substantively different outcomes. With $n_A=2$, both tested initial states reach the proxy corner. Indeed $\alpha_n=0.25$ and $\beta_n=0.6$, so acceptance-oriented production is privately preferable under every reviewer mixture; the careful corner no longer attracts an interior population. Its proxy-corner output is 0.78 value units per researcher-period. This rate result differs from merely changing the value of a paper while retaining the baseline dynamics.

The capacity cap produces a boundary mixture from the lower initial state: $x\simeq0.25$, $y\simeq0$, with accepted value 0.2775. On that edge, unconstrained acceptance demand is $0.65-0.2x$, and the author gap is $0.1-0.06/(0.65-0.2x)$, which vanishes at $x=0.25$. The upper initial state reaches the proxy corner with accepted value 0.18. Thus competition for capacity can replace the pure careful configuration with a mixed author population, even though careful review persists. The affine baseline's corner convergence theorem cannot be transferred to this extension.

Both tested starts remain near their respective attracting corners under the locally matched nonlinear workload functions. Small switching rates instead produce interior states near those corners. At $\mu=0.01$, the lower-start endpoint is approximately $(0.115,0.061)$ and the upper-start endpoint approximately $(0.966,0.949)$. The zero-reviewer-feedback and reversed-author cases reach the careful corner from both tested starts. By contrast, the composition-dependent quality benefit yields residuals of approximately 0.010 and 0.024 at the reported endpoints: these points lie on continuing cycles around the interior center, rather than stationary mixtures. The robustness comparison therefore includes both persistence and failure of the baseline convergence pattern.

\subsection{Verification and reproducibility}
Independent checks compare the saddle Jacobian with finite differences, verify the first integral along an interior trajectory, test state-space bounds, evaluate policy signs on both sides of their thresholds, and confirm the capacity constraint. Zero, reversed, and degenerate feedback cases are included. The policy tests also verify that audits leave the careful-corner author invasion payoff unchanged.

Solver sensitivity is assessed at three declared initial states on a common time grid through time 100. Tightening tolerances from $(10^{-8},10^{-10})$ to $(10^{-10},10^{-12})$ gives a maximum coordinate difference of approximately $1.37\times10^{-7}$ in this comparison. This is a numerical diagnostic, not a bound over every parameter and initial state. Saved configurations, source hashes, package versions, and endpoint CSV files accompany every figure. Numerical definitions and basin tables are generated from these outputs rather than manually transcribed into the manuscript.

The main analysis distinguishes changes in the game from movement within a fixed game. The following experiments make that distinction concrete by varying production costs and reviewer incentives, then examining the state left behind by a temporary intervention.

The cost experiment varies the research-cost difference and the reviewer intercept independently. It fixes $c_A=0.1$ and varies $c_T$ from zero to 0.4, so that $c_T-c_A$ ranges from $-0.1$ to 0.3 without negative production costs. Acceptance probabilities remain at the baseline values. The resulting author intercept is $\alpha=(c_T-c_A)-0.2$, and $\beta=0.4$. The reviewer intercept varies from $-0.4$ to 0.1 while $\kappa=0.4$. Each axis has 41 equally spaced values, and every trajectory starts at $(0.4,0.4)$.

This design separates a production-cost advantage from changes in acceptance preferences. It also separates a reviewer intercept change from changes in workload sensitivity. In the workload interpretation, raising $\gamma$ could reflect an increase in baseline load or a decrease in the private benefit of careful assessment. It is not identified as a particular technology change without additional evidence.

\begin{figure}[tb]\centering\includegraphics[width=\linewidth]{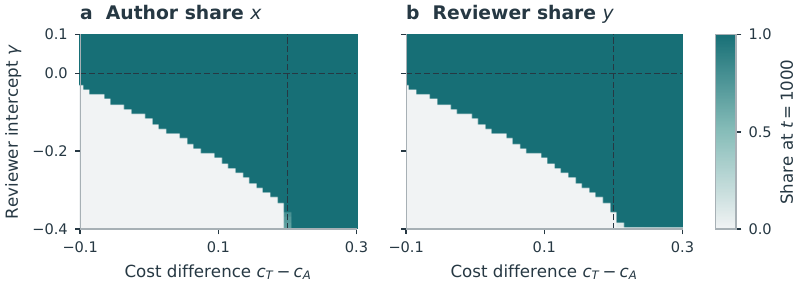}\caption{Finite-time strategy shares under independent changes in research cost difference and reviewer payoff intercept. Both populations start at $(0.4,0.4)$ and are integrated to time 1,000. Dark dashed lines mark $\alpha=0$ and $\gamma=0$, where the careful corner loses strict local stability. A color transition inside the coordination region can instead result from crossing a basin boundary at fixed initial state.}\label{fig:costs}\end{figure}

The panels in Figure~\ref{fig:costs} report endpoints rather than an equilibrium label. This is necessary near $\gamma=0$, $\gamma+\kappa=0$, or an author endpoint equality, where a zero payoff gap can support neutral or slow movement. The strict boundaries are known analytically, so numerical color changes do not supply a separate proof of a bifurcation. The saved cost CSV permits the author and reviewer endpoints to be inspected together.

Within the coordination region, an initial state can move between basins as cost parameters change even before either attracting corner loses stability. This is a form of institutional vulnerability: favorable local incentives near the careful corner do not guarantee that the current population lies in its basin. Conversely, a state already deep in that basin may retain its outcome under a modest cost change. The distinction is between the robustness of a particular trajectory and the local stability of an equilibrium.

\begin{figure}[tb]\centering\includegraphics[width=\linewidth]{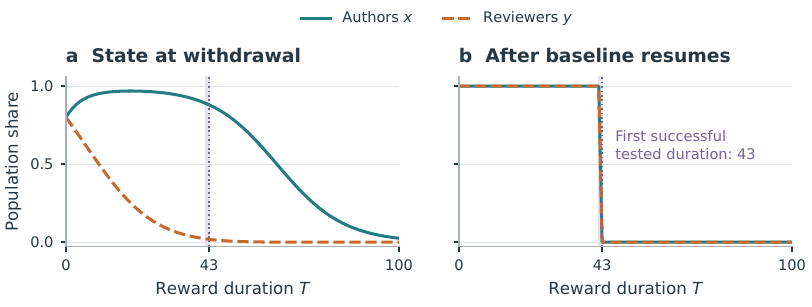}\caption{Temporary reward $s=0.3$ from initial state $(0.8,0.8)$. The left panel shows the state when the reward is withdrawn. The right panel shows the state after a further 4,000 model-time units under baseline incentives. Durations are tested in one-unit increments. The transition identifies a state transfer across the baseline basin boundary at this grid resolution.}\label{fig:temporary}\end{figure}

Figure~\ref{fig:temporary} examines a temporary reward. The release coordinates change continuously with intervention duration, while the asymptotic baseline outcome changes when those coordinates pass the stable manifold. The finite-duration grid finds a first successful duration of 43 model units; durations between 42 and 43 are not resolved by this grid. After release, no subsidy remains in either population's utility. Recovery in this experiment therefore follows from entering the alternative baseline basin, rather than a permanently altered payoff function.

As an independent verification of the basin calculations, the supplementary script solves equation~\eqref{eq:separatrix} on its decreasing branch for every initial state and adaptation speed in the basin CSV. All 5,043 labels agree with the analytical side of that boundary. This checks the basin classification against an implicit analytical characterization rather than merely repeating the same numerical integration at a different tolerance.

\subsection{Supplementary numerical interpretation and diagnostics}
With limited acceptance capacity, careful review need not eliminate acceptance-oriented work. From the lower start, the system approaches $x\simeq0.25$, $y\simeq0$, with accepted value 0.2775. On this edge, unconstrained acceptance demand is $0.65-0.2x$, and the author gap $0.1-0.06/(0.65-0.2x)$ vanishes at $x=0.25$. Proportional thinning reduces the acceptance advantage of exploration until cheaper production offsets it, allowing both author strategies to persist under careful review. From the upper start, proxy assessment persists with accepted value 0.18. This extension shows why improving review effort alone may not resolve the production incentives created by scarce publication slots.

Both tested starts remain near their respective attracting corners under the locally matched nonlinear workload functions. Small switching rates instead produce interior states near those corners. At $\mu=0.01$, the lower-start endpoint is approximately $(0.115,0.061)$ and the upper-start endpoint approximately $(0.966,0.949)$. The zero-reviewer-feedback and reversed-author cases reach the careful corner from both tested starts. By contrast, the composition-dependent quality benefit yields residuals of approximately 0.010 and 0.024 at the reported endpoints: these points lie on continuing cycles around the interior center, rather than stationary mixtures. The robustness comparison therefore includes both persistence and failure of the baseline convergence pattern.

\section{Evidence and reproducibility}\label{app:evidence}
The evidence and numerical records are retained here to separate empirical motivation from model calibration and implementation checks.
\label{sec:evidence}
The literature suggests a plausible feedback, but its components have different evidential support. This section examines scale, evaluation practices, and selection incentives in turn, then identifies the measurements still needed to connect them. The purpose is to ground the model in observed mechanisms rather than to treat its payoff signs as empirical facts.

\subsection{Scale and evaluation demands}
The official ICLR 2026 review retrospective records 19,525 valid, format-compliant submissions and 76,139 reviews by 18,054 reviewers. It distinguishes this submission count from the 13,763 papers receiving acceptance or rejection decisions after withdrawals and desk rejections \citep{iclr2026}. These counts illustrate the scale of evaluation activity. They do not establish that automation caused submission growth or that a particular reviewer experienced an increase in assigned load.

Even official records can use different reporting conventions. The retrospective reports 5,355 acceptances, whereas the conference fact sheet reports 5,357 and lists 21,674 reviewers \citep{iclr2026facts}. We retain the retrospective's internally described process counts and do not combine its active-reviewer denominator with the fact sheet's reviewer count. The two-acceptance discrepancy remains a reporting difference rather than an adjustment to our data. No acceptance-rate estimate from these records enters the game parameters.

Workload in the model is an opportunity-cost concept. The relevant burden includes time to understand unfamiliar work, check claims, inspect artifacts, and resolve discrepancies. Aggregate review counts motivate measuring that burden, rather than measuring its marginal response to author strategy. A proper estimate would require assignments, reviewer time, verification tasks, and submission characteristics.

\subsection{Assistance and delegated judgment}
\citet{liang2024} estimate that 6.5--16.9\% of review text in the studied conference corpora could have been substantially modified by LLMs. Their estimates concern text at the corpus level. They do not count the proportion of reviewers who delegated their scientific judgment. The same writing pattern can arise from polishing an independently formed assessment or from generating an assessment with little independent checking.

\begin{figure}[tb]\centering\includegraphics[width=\linewidth]{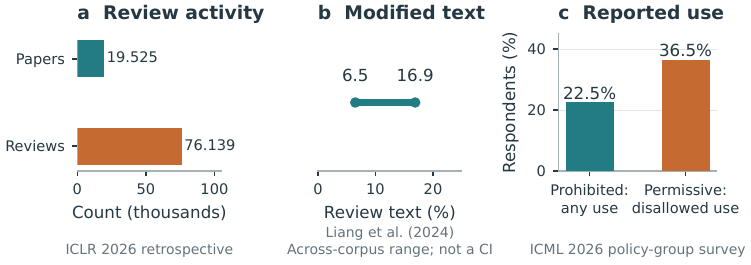}\caption{Separate empirical indicators. Left: process counts from the ICLR 2026 retrospective. Middle: estimated substantially LLM-modified review text across the conference corpora studied by Liang et al. (2024); the range is not a confidence interval. Right: self-reported use among respondents in two ICML policy groups, with distinct definitions. The panels do not share a denominator and do not calibrate the game.}\label{fig:evidence}\end{figure}

\citet{kim2026} report a randomized policy experiment and a post-survey with 1,486 responses at ICML 2026. Among respondents assigned the conservative policy, 22.5\% reported some LLM use; among those assigned the permissive policy, 36.5\% reported at least one disallowed use. Policy assignment had near-zero effects on final scores, decisions, and reviewer confidence. These results show why actual evaluation practices cannot be inferred from policy labels. The survey percentages have different behavioral definitions and group denominators; they are displayed separately in Figure~\ref{fig:evidence}.

The interpretation is that low-cost assistance is available and used, including under constraints. It remains necessary to measure when assistance substitutes for costly checking, when it supports checking, and how the cost difference responds to workload. A rule permitting language assistance is not equivalent to an increase in the population share $y$.

\subsection{Autonomous production and review failures}
The direct production evidence concerns the reliability of completed research, not merely its writing. \citet{beel2025evaluation} report five execution failures among twelve experiments and incorrect or hallucinated numbers in four of seven resulting manuscripts. \citet{luo2026pitfalls} identify hidden methodological failures in two systems; their balanced audit dataset includes naturally occurring and induced cases. Manuscript-only audit accuracy is 53.7\%, compared with 81.5\% with workflow artifacts. SciIntegrity-Bench reports a combined failure-and-flawed-output rate of 34.2\% across 231 runs in deliberately constructed integrity dilemmas \citep{yang2026integrity}. These results expose ways in which completing a paper can become detached from establishing a sound contribution.

The review evidence supplies the complementary vulnerability. BadScientist's fabricated submissions reach acceptance rates of up to 82\% in tested adversarial configurations \citep{jiang2026badscientist}; this is an AI-review outcome, not an observed conference acceptance rate. Content-preserving rhetorical changes also move AI review scores \citep{li2026rhetoric}. Review-focus comparisons and the Austrian grant study identify differences in attention to technical validity and novelty \citep{shin2025blindspots,machado2026novelty}. Together, these findings motivate the possibility that reviewer-compatible presentation receives rewards that uncertain scientific promise does not. The game examines how that gap can persist through adaptation.

\subsection{Evaluation incentives and scientific value}
Novelty-related evaluation difficulties predate current AI systems. In a grant-selection setting, \citet{boudreau2016} randomized evaluator--proposal assignments and found systematic associations of scores with novelty and intellectual distance. This supports treating the recognition of exploratory work as an institutional problem shared by human and automated evaluation. It does not establish that careful humans always favor exploratory work or that the cross-protocol coefficient $\beta$ is positive.

\citet{sharma2026} analyze over 125,000 paper--review pairs. An apparent preference of LLM-associated reviews for LLM-associated papers is explained by quality differences in their analysis. Their fully generated reviews exhibit compressed ratings and weaker quality discrimination, while human reviewers using LLMs reduce this leniency. The feedback mechanism does not require an inherent preference for AI authors: weak discrimination can reward inexpensive, evaluator-compatible work regardless of who produced it.

Evidence from funding provides an adjacent comparison. \citet{qian2026} combine confidential NSF and NIH submissions from two universities with public award records. Greater inferred LLM involvement is associated with lower semantic distinctiveness. Associations with funding success and early publication output differ by agency, with positive associations reported at NIH and no comparable associations at NSF. The reported NIH publication gains are concentrated outside the most highly cited papers. These are observational relationships in grant funding; semantic distinctiveness and citations do not directly measure the scientific value in our welfare function.

Together, these observations motivate the warning that evaluation rewards can become detached from contribution. The feedback coefficient $\beta$ concerns the relative rewards of research strategies under different protocols; estimating it requires observing those strategies and outcomes together.

\begin{table}[tb]
\caption{From observations to model quantities. The final column identifies the measurement still needed.}\label{tab:evidence}\small
\begin{tabularx}{\linewidth}{@{}p{.27\linewidth}p{.20\linewidth}X@{}}\toprule
Observation & Model connection & Additional measurement\\\midrule
Large review volume & Load $L(x)$ & Time and verification burden conditional on submission traits\\
LLM-associated review text & Availability of $M$ & Delegated judgment versus assisted checking\\
Protocol-dependent discrimination & Interaction $\beta$ & Relative reward changes holding content and validity comparable\\
Policy noncompliance & Incentive response & Behavioral response to review costs and rewards\\
Selection versus distinctiveness & Value $v_i$ & Independent validation with a stated horizon and unit\\\bottomrule
\end{tabularx}\end{table}

Section~\ref{sec:model} treats the payoff coefficients as unknown institutional quantities. The evidence record distributed with the code preserves source versions, definitions, and limitations. No individual is classified with an AI-text detector, and no published adoption percentage is substituted for an evolutionary state.

The empirical sources were checked on November 7, 2026. The evidence record includes the review-process retrospective, conference fact sheet, review-text study, paper--review interaction study, policy experiment, and funding study. The closest theoretical and synthesis papers are included for positioning. Each record states a version or publication date, population, measurement or finding, uncertainty interpretation, and limitation.

For administrative records, the main issue is category definition and reporting date rather than a sampling confidence interval. For the modified-text study, the reported range spans corpus estimates; it should not be plotted as an uncertainty interval around a common population mean. For the policy survey, the two quoted percentages refer to different forms of behavior within different assigned groups. A pooled percentage would require additional counts and an explicit estimand.

The observational funding findings concern inferred text involvement and agency-specific associations. They motivate separating rewards from scientific distinctiveness, but no numerical coefficient is imported into $v_A$, $v_T$, or the production rates. Confidential rejected proposals are not part of the reproducibility package. The generated empirical figure contains only source-level aggregates and does not attempt individual attribution.

The empirical studies identify vulnerabilities in research production and evaluation. The model connects them through incentives, showing how repeated adaptation can turn individual failures into a persistent institutional problem. A direct empirical test would jointly measure the protocol interaction, workload response, and independently assessed scientific contribution.

\subsection{Extended literature and empirical agenda}
The feedback described in the introduction connects three bodies of work: incentive-driven changes in scientific practice, empirical studies of evaluation, and formal models of adaptation. We use these literatures to locate the mechanism and to distinguish established results from the questions addressed here.

\subsubsection{Incentives, novelty, and research practices}
\citet{holmstrom1991} explain why incentives attached to measurable tasks can divert effort from less measurable activities. Applied to scientific work, this supplies an institutional interpretation of optimizing acceptance signals while underinvesting in difficult verification. \citet{smaldino2016} model how publication-based selection can propagate weaker research methods without deliberate misconduct. Our model adds an adapting evaluator population and isolates the conditions under which its choices reinforce author behavior.

The novelty dimension has an empirical precedent in \citet{boudreau2016}, who study evaluations of frontier research proposals under randomized evaluator assignment. Their results make expertise and novelty relevant to evaluation design, while leaving open the human--AI protocol interaction in our game. \citet{messeri2024} offer a conceptual account of epistemic risks from AI-mediated science. We use that perspective to motivate an independently assessed scientific objective, rather than to supply a numerical welfare coefficient.

\subsubsection{Verification incentives and adaptation}
The closest theoretical comparison is \citet{you2026}. Their model links verification capacity and proxy noise to the correlation between evaluation scores and latent scientific value. They also study how a researcher allocates effort between truth-oriented and proxy-oriented activities. Our model instead treats author and reviewer strategy shares as two evolving states. This permits separate questions about reviewer adoption, equilibrium selection, and intervention withdrawal. We use neither their score-correlation formula nor their effort-production function. The common institutional concern is that evaluation rewards can become weakly connected to the value of research.

\citet{wang2026} place production, evaluation, manipulation, institutional response, and longer-run feedback within a broader account of scholarly publishing. Their synthesis is an important reason to frame our contribution narrowly. We formalize one behavioral mechanism within that larger system: the interaction of acceptance incentives with workload-dependent review effort. The model does not represent every step in their taxonomy. In particular, strategic evasion, persistent scholarly records, and training future systems on generated material are outside the baseline dynamics.

These comparisons distinguish three levels of explanation. A score model asks how accurately evaluation reflects a contribution. An individual choice model asks how a researcher responds to that evaluation. A population model asks how many participants adopt each practice and how those practices change one another's payoffs. The third level is our focus. It can support path dependence even when the evaluation technology itself is fixed, because the population state changes the environment in which that technology is used.

\subsubsection{Automated assistance and evaluation behavior}
Corpus-level work by \citet{liang2024} documents the entry of low-cost language-model assistance into conference reviewing. For the feedback mechanism, the consequential change is reduced independent checking under review pressure. Measuring verification actions alongside tool use would establish whether assistance strengthens checking or substitutes for it.

\citet{sharma2026} examine interactions between LLM-associated papers and reviews, including how apparent preferences change after quality adjustment. Our author payoff interaction asks whether a review protocol changes the relative rewards of two research strategies. Testing that interaction requires observing research characteristics and evaluation outcomes together. It cannot be inferred from authorship labels alone.

The policy experiment of \citet{kim2026} further separates permission, reported use, and review outcomes. Its randomized contrast concerns assigned rules for assistance, rather than a random assignment to fully automated scientific judgment. This distinction matters when interpreting a null policy effect. It also suggests that incentives and actual behavior should be measured alongside written policy when evaluating an institutional intervention.

\subsubsection{Institutional alternatives}
Institutional design can change evaluation supply as well as evaluation rules. \citet{stelmakh2020} report a conference experiment combining recruitment and guidance of novice reviewers, with favorable review-quality assessments relative to the conventional pool. This motivates reviewer development as a way to alter workload and effective capacity. \citet{goldberg2024} examine feedback on reviews in a randomized trial and related experiments, also documenting substantial disagreement about review quality. The latter finding cautions against equating a single review-quality score with a reliable reward signal.

Registered Reports move part of evaluation before the outcome of a study is known \citep{chambers2013,chambers2022}. Their relevance here is a change in what earns an initial publication commitment, not a presumption that all exploratory research can be preregistered. The transparency standards proposed by \citet{nosek2015} offer complementary ways to make evidence inspectable. These approaches motivate the design proposals in Section~\ref{sec:discussion}; their ability to reverse this game's dynamics remains a model-based hypothesis.

\subsubsection{Evolutionary interpretation}
Replicator dynamics describe how strategies with higher payoffs increase their population shares. Their use here follows established evolutionary game theory \citep{hofbauer1998}. The equations can represent imitation, differential persistence, or gradual changes in the allocation of activity. We do not interpret them as a literal biological process or assume that academic behavior follows a known adjustment rate.

The coordination structure and its elementary stability properties are established mathematical tools. Their value in this setting lies in the interpretation of the coefficients and in the separation between private rewards and scientific welfare. Explicit thresholds make it possible to ask what an audit must change, which population a subsidy affects, and why removing a policy can have different consequences depending on the state reached before withdrawal. These are the questions that the subsequent analysis addresses.

\subsection{Empirical tests}

A direct test would estimate the acceptance interaction $\beta$ by comparing research strategies across review protocols. Research content, technical soundness, and presentation should be distinguished. Rewriting a fixed paper can identify sensitivity to presentation, but it cannot by itself establish that exploratory research has a different production value or cost. A useful design would combine controlled presentation changes with independently assessed research characteristics and reviewer effort measurements.

The workload feedback requires another design. Assignments or time constraints could vary independently of paper characteristics, allowing measurement of whether additional load changes actual checking and delegation. Reported tool use alone is insufficient. Observable verification actions, time spent on consequential claims, and the quality of resulting evidence would be more informative. Changes in reviewer composition and expertise would need to be distinguished from within-reviewer responses.

Scientific value is the hardest quantity. For some computational claims, held-out performance, independent replication, proof verification, or discovery in a hidden environment can provide a useful outcome. For conceptual work, value may require longer horizons and several forms of expert judgment. Citations and acceptance decisions are relevant institutional outcomes, but using them as the sole value measure risks reproducing the proxy problem being studied. A future empirical version should declare its value measure and horizon before fitting the welfare comparison.

The empirical tests should establish whether proxy evaluation raises the relative reward of acceptance-oriented work, whether production pressure raises the return to proxy assessment, and whether the resulting output loses scientific value. These are the three links through which the modeled cycle damages the community. Zero or reversed feedback and sufficiently valuable output mark boundaries where that mechanism does not produce the same loss.

\subsection{Model boundaries}
The baseline abstracts from heterogeneity in expertise, preferences, and access to tools. Real research practices form a continuum, and participants often mix exploratory work with conventional contributions. Binary strategies expose the coordination mechanism but do not estimate how sharply those practices divide. Extending the strategy space would require additional behavioral parameters rather than a relabeling of the same two types.

The acceptance probabilities summarize a whole evaluation institution. Multiple reviewers, area chairs, rebuttals, and correlated errors can generate nonlinear decision rules. A conference with fixed capacity also creates competitive effects that are absent from independent acceptance probabilities. The capacity sensitivity case addresses one consequence but uses proportional thinning rather than a realistic merit-ranking process. Its findings therefore describe that specified mechanism.

The reviewer workload equation is deliberately transparent. Its empirical sign can change when automation reduces verification time, when reviewer supply grows with submissions, or when acceptance-oriented papers are easier to check. Accuracy benefits can also change with paper composition. The nonlinear and reversed-feedback experiments show how the proposed reinforcement can fail, but they do not exhaust possible institutional responses.

Finally, the adjustment dynamics omit strategic foresight, explicit coordination, and shocks. Exact boundaries are absorbing without switching, whereas real communities continually admit new participants and practices. A deterministic basin diagram should be read as a conditional map of incentives. It does not establish the location of an actual conference in that map, the probability of a shock, or a calendar date at which a transition might occur.

%% file: figures/basin_table.tex
\begin{tabular}{lrrr}
\toprule
$\rho$ & Careful & Proxy & Unresolved / saddle\\
\midrule
0.25 & 462 & 1219 & 0\\
1 & 532 & 1149 & 0\\
4 & 642 & 1039 & 0\\
\bottomrule
\end{tabular}